\documentclass[a4paper,12pt]{article}
\usepackage{amssymb}
\usepackage{amsmath}
\usepackage{fullpage}
\usepackage{boxedminipage}
\usepackage{listings}
\usepackage{minitoc}

\makeatletter \@addtoreset{equation}{section} \makeatother

\begin{document}

%\ifpdf\DeclareGraphicsExtensions{.pdf, .jpg, .tif} \else%
%\DeclareGraphicsExtensions{.eps, .jpg} \fi
\begin{titlepage}

    \thispagestyle{empty}
    \begin{flushright}
        \hfill{CERN-PH-TH/2008-090}\\\hfill{UCLA/08/TEP/12}
    \end{flushright}

    \vspace{5pt}
    \begin{center}
        { \Huge{\textbf{Attractors in Black}}}\vspace{25pt}
%\\
%\textit{\Large{\textbf{Extremal Black Holes and
%Attractors\\\vspace{7pt}in $\mathcal{N}$=2, d=4 Maxwell-Einstein
%Supergravity}}}
        \vspace{30pt}

        {\bf Stefano Bellucci$^\clubsuit$, Sergio Ferrara$^{\diamondsuit\clubsuit\flat}$ and\ Alessio Marrani$^{\heartsuit\clubsuit}$}

        \vspace{15pt}

        {$\clubsuit$ \it INFN - Laboratori Nazionali di Frascati, \\
        Via Enrico Fermi 40,00044 Frascati, Italy\\
        \texttt{bellucci,marrani@lnf.infn.it}}

        \vspace{10pt}

        {$\diamondsuit$ \it Physics Department,Theory Unit, CERN, \\
        CH 1211, Geneva 23, Switzerland\\
        \texttt{sergio.ferrara@cern.ch}}

        \vspace{10pt}
         {$\flat$ \it Department of Physics and Astronomy,\\
        University of California, Los Angeles, CA USA\\
        \texttt{ferrara@physics.ucla.edu}}

         \vspace{10pt}

        {$\heartsuit$ \it Museo Storico della Fisica e\\
        Centro Studi e Ricerche ``Enrico Fermi"\\
        Via Panisperna 89A, 00184 Roma, Italy}

        \vspace{30pt}
        \noindent \textit{Contribution to the Proceedings of the 3rd RTN Workshop\\``Constituents, Fundamental Forces and Symmetries of the Universe'',\\1--5 October 2007, Valencia, Spain}
\end{center}

%\vspace{50pt}

\begin{abstract}

We review recent results in the study of attractor horizon
geometries (with non-vanishing Bekenstein-Hawking entropy) of dyonic
extremal $d=4$
black holes in supergravity. We focus on $%
\mathcal{N}=2$, $d=4$ ungauged supergravity coupled to a number
$n_{V}$ of
Abelian vector multiplets, outlining the fundamentals of the special K\"{a}%
hler geometry of the vector multiplets' scalar manifold (of complex
dimension $n_{V}$), and studying the $\frac{1}{2}$-BPS attractors,
as well as the non-BPS (non-supersymmetric) ones with non-vanishing
central charge.

For symmetric special K\"{a}hler geometries, we present the complete
classification of the orbits in the symplectic representation of the classical $U$%
-duality group (spanned by the black hole charge configuration
supporting the attractors), as well as of the moduli spaces of
non-BPS attractors (spanned by the scalars which are not stabilized
at the black hole event horizon).

Finally, we report on an analogous classification for $\mathcal{N}>2$%
-extended, $d=4$ ungauged supergravities, in which also the $\frac{1}{%
\mathcal{N}}$-BPS attractors yield a related moduli space.
\end{abstract}

\end{titlepage}
\newpage\tableofcontents%\baselineskip6 mm

\section{Introduction\label{Intro}}

\textit{Extremal }black hole (BH) \textit{attractors} were
discovered some time ago in
\cite{FKS}-\nocite{Strom,FK1,FK2}\cite{FGK}. Recently, a number of
papers have been devoted to their study \cite{Sen-old1}--\nocite
{GIJT,Sen-old2,K1,TT,G,GJMT,Ebra1,K2,Ira1,Tom,
BFM,AoB-book,FKlast,Ebra2,bellucci1,rotating-attr,K2-bis,Misra1,Lust2,Morales,BFMY,Astefa,CdWMa,
DFT07-1,BFM-SIGRAV06,Cer-Dal,ADFT-2,Saraikin-Vafa-1,Ferrara-Marrani-1,TT2,
ADOT-1,ferrara4,CCDOP,Misra2,Astefanesei,Anber,Myung1,Ceresole,BMOS-1,Hotta,
Gao,PASCOS07,Sen-review,Belhaj1,AFMT1,Gaiotto1,BFMS1,GLS1,ANYY1,review-Kallosh,Cai-Pang,Vaula,
Li,BFMY2,Saidi2,Saidi3}\cite{Saidi4} (for further developments, see
also \textit{e.g.}
\cite{OSV}--\nocite{OVV,ANV,GSV}\cite{Pioline-review}), essentially
because new classes of solutions to the so-called \textit{Attractor
Equations} were
(re)discovered. Such new solutions have been found to determine non-BPS (%
\textit{Bogomol'ny-Prasad-Sommerfeld}) BH horizon geometries, breaking
\textit{all} supersymmetries (\textit{if any}).

The near-horizon attractor geometry of an extremal (static, spherically
symmetric, asymptotically flat, dyonic) BH is associated to the
corresponding configuration of the $1\times \left( 2n_{V}+2\right) $
symplectic vector of the BH magnetic and electric charges $Q\equiv \left(
p^{\Lambda },q_{\Lambda }\right) $, defined as the spatially asymptotical
fluxes of the vector field-strengths:
\begin{equation}
~p^{\Lambda }\equiv \frac{1}{4\pi }\int_{S_{\infty }^{2}}\mathcal{F}%
^{\Lambda },~~~q_{\Lambda }\equiv \frac{1}{4\pi }\int_{S_{\infty }^{2}}%
\mathcal{G}_{\Lambda }.  \label{Gamma-tilde}
\end{equation}
The symplectic index $\Lambda $ run $0,1,...,n_{V}$. In $\mathcal{N}=2$, $%
d=4 $ \textit{ungauged} supergravity $n_{V}$ denotes the number of Abelian
vector multiplets coupled to the supergravity one\footnote{%
The \textit{Attractor Mechanism} in $\mathcal{N}=2$, $d=4$ \textit{ungauged}
supergravity does \textit{not} deal with the $n_{H}$ hypermultiplets
eventually present.
\par
This is ultimately due to the transformation properties of the Fermi fields:
the \textit{hyperinos} $\zeta ^{\alpha }$ transform independently on the
vector fields, whereas the supersymmetry transformations of the \textit{%
gauginos} $\lambda ^{i}$ do depend on the Maxwell vector fields (see \textit{%
e.g.} \cite{Castellani1,DFF}). Consequently, the contribution of the
hypermultiplets may be dynamically decoupled from the rest of the physical
system. Thus, it is also completely independent from the evolution dynamics
of the vector multiplets' scalars.} (which contains the Maxwell vector $%
A^{0} $, usually named \textit{graviphoton}). Moreover, denoting with $r$
the radial coordinate, $S_{\infty }^{2}$ is the $2$-sphere for $r\rightarrow
\infty $. $\mathcal{F}^{\Lambda }=dA^{\Lambda }$ and $\mathcal{G}_{\Lambda }$
is the related \textit{``dual''} field-strength two-form (see \textit{e.g.}
\cite{3,4}; see also \cite{Lauwers,Strom-1}).

The present report will deal only with \textit{non-degenerate }($\frac{1}{2}$%
-BPS as well as non-BPS) geometries, \textit{i.e.} with geometries yielding
a finite,\textit{\ non-vanishing} (effective) horizon area $A_{H}$,
corresponding to the so-called \textit{``large''} BHs. Through the \textit{%
critical implementation} of the so-called \textit{Attractor Mechanism} \cite
{FKS}-\nocite{Strom,FK1,FK2}\cite{FGK}, the \textit{Bekenstein-Hawking
entropy} $S_{BH}$ \cite{BH1} associated to such attractor geometries\textbf{%
\ }can be computed by extremizing a properly defined, positive-definite
\textit{``effective BH potential''} $V_{BH}\left( \phi ,p,q\right) $, with ``%
$\phi $'' standing for the relevant set of real scalar fields.

In $\mathcal{N}=2$, $d=4$ supergravity non-degenerate attractor horizon
geometries split in two classes, one corresponding to $\frac{1}{2}$-BPS
\textit{``short massive multiplets''} (preserving four supersymmetries out
of the eight pertaining to the $\mathcal{N}=2$, $d=4$ superPoincar\'{e}
asymptotical background), and the other given by non-supersymmetric \textit{%
``long massive multiplets''} violating the BPS bound \cite{BPS}:
\begin{equation}
\begin{array}{l}
\frac{1}{2}\text{\textit{-BPS}:~~}0<\left| Z\right| _{H}^{2}=S_{BH}/\pi ; \\
~ \\
\text{\textit{non-BPS~,}}\left\{
\begin{array}{l}
Z\neq 0\text{:~~}0<\left| Z\right| _{H}^{2}<S_{BH}/\pi ; \\
\\
Z=0\text{:~~}0=\left| Z\right| _{H}^{2}<S_{BH}/\pi ,
\end{array}
\right. \text{ }
\end{array}
\end{equation}
where the subscript ``$H$'' denotes the evaluation at the BH event horizon,
and $Z$ stands for the $\mathcal{N}=2$, $d=4$ \textit{central charge
function }(see \textit{e.g.} \cite{4} and Refs. therein). As mentioned, the
\textit{Bekenstein-Hawking entropy} $S_{BH}$ is obtained by extremizing $%
V_{BH}\left( \phi ,p,q\right) $ with respect to its dependence on the
scalars \cite{BH1,FGK}:
\begin{equation}
S_{BH}\left( p,q\right) =\frac{A_{H}\left( p,q\right) }{4}=\pi \left[
V_{BH}\left( \phi ,p,q\right) \right] _{\partial _{\phi }V_{BH}=0}=\pi
V_{BH}\left( \phi _{H}\left( p,q\right) ,p,q\right) .  \label{BHEA}
\end{equation}
The purely \textit{charge-dependent} horizon configuration $\phi _{H}\left(
p,q\right) $ of the real scalars is a solution of the \textit{criticality
conditions}
\begin{equation}
\frac{\partial V_{BH}\left( \phi ,p,q\right) }{\partial \phi }=0,
\label{crit-cond-gen}
\end{equation}
and it determines an attractor in a strict sense if the critical $\left(
2n_{V}+2\right) \times \left( 2n_{V}+2\right) $ real symmetric Hessian
matrix
\begin{equation}
\mathcal{H}^{V_{BH}}\equiv \left. \frac{\partial ^{2}V_{BH}\left( \phi
,p,q\right) }{\partial \phi \partial \phi }\right| _{\phi =\phi _{H}\left(
p,q\right) }  \label{crit-Hessian-gen}
\end{equation}
is \textit{strictly positive-definite}.

It should be remarked that the opposite does \textit{not} hold in general,
\textit{i.e.} attractors may exist such that the corresponding $\mathcal{H}%
^{V_{BH}}$ exhibits some vanishing eigenvalues. If this happens, a careful
study of the (signs of the) higher-order covariant derivatives of $V_{BH}$
evaluated at the considered critical point(s) is needed. Depending on the
supporting BH charge configuration, the \textit{massless Hessian modes} can
be lifted to positive values (determining \textit{stable} critical points,
and thus attractors) or to negative values (yielding \textit{unstable}
critical points, and thus \textit{repellers}). Examples in \ literature of
investigations beyond the Hessian level can be found in \cite
{TT,K2-bis,Misra1}. But a third possibility may happen, namely that the
massless Hessian modes persist at \textit{all} order in the covariant
differentiation of $V_{BH}$; in such a case, a \textit{moduli space} arises
out, spanned by the scalars which are \textit{not} stabilized at the BH
event horizon in terms of the BH charges belonging to the configuration
supporting the considered class of critical points of $V_{BH}$.

Non-supersymmetric (non-BPS) BH attractors arise also in $\mathcal{N}>2$%
-extended, $d=4$ and $d=5$ supergravities \cite{FG,FKlast} (see \textit{e.g.}
\cite{ADFT,review-Kallosh,Larsen-review,d=5-review} for recent reviews), but
$\mathcal{N}=2$, $d=4$ supergravity, whose scalar manifold is a special
K\"{a}hler (SK) space, exhibits the richest case study.

Moduli spaces of attractors have been recently found and classified in \cite
{Ferrara-Marrani-1,ferrara4} for $\mathcal{N}=2$ \textit{symmetric} and $%
\mathcal{N}>2$-extended, $d=4$ supergravities (see also \cite{GLS1} for an
explicit treatment in the so-called $stu$ model). In such theories, the
Hessian matrix of $V_{BH}$ at its critical points is in general positive
definite, eventually with some vanishing eigenvalues, which actually are
\textit{flat} directions of $V_{BH}$ itself. More in general, it can be
stated that for \textit{all} supergravities based on \textit{homogeneous}
(not necessarily symmetric) scalar manifolds the \textit{non-degenerate }%
critical points of $V_{BH}$ are \textit{all} \textit{stable}, up to some
eventual \textit{flat} directions. We will briefly report on such an issue
in the last Section.\bigskip

The plan of the paper is as follows.

Sect. \ref{N=2,d=4} reports about extremal BH attractors in $\mathcal{N}=2$,
$d=4$ supergravity. The fundamentals of the SK geometry of the scalar
manifold are outlined in Subsect. \ref{SKG-gen}. Thence, in Subsect. \ref
{Sect2} $V_{BH}$ for $\mathcal{N}=2$, $d=4$ supergravity is introduced, and
its $\frac{1}{2}$-BPS critical points \cite{FKS}- \nocite{Strom,FK1,FK2}\cite
{FGK}, which are always \textit{stable} (thus determining attractors in a
strict sense), are studied. The features of the class of non-BPS critical
points of $V_{BH}$ with non-vanishing $Z$ are presented in Subsect. \ref
{Sect3}. The class of non-BPS critical points of $V_{BH}$ with $Z=0$ will
not be considered here (see rather \textit{e.g.} \cite{review-Kallosh} and
Refs. therein). Sect. \ref{Conclusion} reports some recent results on the
classification of the supporting BH \textit{charge orbits} and \textit{%
moduli spaces} of extremal BH attractors in $\mathcal{N}=2$ \textit{symmetric%
} (Subsect. \ref{N=2-d=4-symm}) and $\mathcal{N}>2$-extended (Subsect. \ref
{N>2,d=4}), $d=4$ supergravities.\setcounter{equation}0

\section{\label{N=2,d=4}Extremal Black Hole Attractors\newline
in $\mathcal{N}=2$, $d=4$ Supergravity}

\subsection{\label{SKG-gen}\textit{Glossary} of Special K\"{a}hler Geometry}

In the present Section we briefly recall the fundamentals of the SK geometry
underlying the scalar manifold $\mathcal{M}_{n_{V}}$ of $\mathcal{N}=2$, $%
d=4 $ supergravity coupled to $n_{V}$ Abelian vector multiplets ($dim_{%
\mathbb{C}}\mathcal{M}_{n_{V}}=n_{V}$; see \cite{Lauwers,Strom-1}).

It is convenient to switch from the Riemannian $2n_{V}$-dim. parametrization
of $\mathcal{M}_{n_{V}}$ given by the local real coordinates $\left\{ \phi
^{a}\right\} _{a=1,...,2n_{V}}$ to the K\"{a}hler $n_{V}$-dim.
holomorphic/antiholomorphic parametrization given by the local complex
coordinates $\left\{ z^{i},\overline{z}^{\overline{i}}\right\} _{i,\overline{%
i}=1,...,n_{V}}$. This corresponds to the performing the \textit{unitary
Cayley transformation}:
\begin{equation}
z^{k}\equiv \frac{\varphi ^{2k-1}+i\varphi ^{2k}}{\sqrt{2}},~~k=1,...,n_{V}.
\label{unit-transf}
\end{equation}
The metric structure of $\mathcal{M}_{n_{V}}$ is given by the covariant SK
metric tensor\footnote{%
Usually, the $n_{V}\times n_{V}$ Hermitian matrix $g_{i\overline{j}}$ is
assumed to be invertible, with non-vanishing determinant and rank $n_{V}$,
and with Euclidean signature (\textit{i.e.} with all strictly positive
eigenvalues)\textit{\ globally} in $\mathcal{M}_{n_{V}}$. We will so assume,
even though we will be concerned mainly with the properties of $g_{i%
\overline{j}}$ at those peculiar points of $\mathcal{M}_{n_{V}}$ which are
critical points of $V_{BH}$.} $g_{i\overline{j}}\left( z,\overline{z}\right)
=\partial _{i}\overline{\partial }_{\overline{j}}K\left( z,\overline{z}%
\right) $, $K\left( z,\overline{z}\right) $ being the real K\"{a}hler
potential.

The previously mentioned $\mathcal{N}=2$, $d=4$ \textit{central charge
function} is defined as (see \textit{e.g.} \cite{4} and refs. therein)
\begin{eqnarray}
Z\left( z,\overline{z};q,p\right) &\equiv &Q\Omega V\left( z,\overline{z}%
\right) =q_{\Lambda }L^{\Lambda }\left( z,\overline{z}\right) -p^{\Lambda
}M_{\Lambda }\left( z,\overline{z}\right) =e^{\frac{1}{2}K\left( z,\overline{%
z}\right) }Q\Omega \Pi \left( z\right) =  \notag \\
&&  \notag \\
&=&e^{\frac{1}{2}K\left( z,\overline{z}\right) }\left[ q_{\Lambda
}X^{\Lambda }\left( z\right) -p^{\Lambda }F_{\Lambda }\left( z\right) \right]
\equiv e^{\frac{1}{2}K\left( z,\overline{z}\right) }W\left( z;q,p\right) ,
\label{Z}
\end{eqnarray}
where $\Omega $ is the $\left( 2n_{V}+2\right) $-dim. square symplectic
metric (subscripts denote dimensions of square sub-blocks)
\begin{equation}
\Omega \equiv \left(
\begin{array}{ccc}
0_{n_{V}+1} &  & -\mathbb{I}_{n_{V}+1} \\
&  &  \\
\mathbb{I}_{n_{V}+1} &  & 0_{n_{V}+1}
\end{array}
\right) ,  \label{Omega}
\end{equation}
and $V\left( z,\overline{z}\right) $ and $\Pi \left( z\right) $ respectively
stand for the $\left( 2n_{V}+2\right) \times 1$ covariantly holomorphic
(K\"{a}hler weights $\left( 1,-1\right) $) and holomorphic (K\"{a}hler
weights $\left( 2,0\right) $) period vectors in symplectic basis:
\begin{gather}
\overline{D}\overline{_{i}}V\left( z,\overline{z}\right) =\left( \overline{%
\partial }_{\overline{i}}-\frac{1}{2}\overline{\partial }_{\overline{i}%
}K\right) V\left( z,\overline{z}\right) =0,~~~D_{i}V\left( z,\overline{z}%
\right) =\left( \partial _{i}+\frac{1}{2}\partial _{i}K\right) V\left( z,%
\overline{z}\right) ;  \notag \\
\Updownarrow  \notag \\
V\left( z,\overline{z}\right) =e^{\frac{1}{2}K\left( z,\overline{z}\right)
}\Pi \left( z\right) ,~~\overline{D}\overline{_{i}}\Pi \left( z\right) =%
\overline{\partial }_{\overline{i}}\Pi \left( z\right) =0,~~~D_{i}\Pi \left(
z\right) =\left( \partial _{i}+\partial _{i}K\right) \Pi \left( z\right) ;
\notag \\
\notag \\
\Pi \left( z\right) \equiv \left(
\begin{array}{c}
X^{\Lambda }\left( z\right) \\
\\
F_{\Lambda }\left( X\left( z\right) \right)
\end{array}
\right) =exp\left( -\frac{1}{2}K\left( z,\overline{z}\right) \right) \left(
\begin{array}{c}
L^{\Lambda }\left( z,\overline{z}\right) \\
\\
M_{\Lambda }\left( z,\overline{z}\right)
\end{array}
\right) ,  \label{PI}
\end{gather}
with $X^{\Lambda }\left( z\right) $ and $F_{\Lambda }\left( X\left( z\right)
\right) $ being the holomorphic sections of the $U(1)$ line (Hodge) bundle
over $\mathcal{M}_{n_{V}}$. $W\left( z;q,p\right) $ is the so-called \textit{%
holomorphic }$\mathcal{N}=2$\textit{\ central charge function}, also named $%
\mathcal{N}=2$ \textit{superpotential} ($\overline{\partial }_{\overline{i}%
}W=0$).

Up to some particular choices of local symplectic coordinates in $\mathcal{M}%
_{n_{V}}$, the covariant symplectic holomorphic sections $F_{\Lambda }\left(
X\left( z\right) \right) $ may be seen as derivatives of an \textit{%
holomorphic prepotential} function $F$ (with K\"{a}hler weights $\left(
4,0\right) $):
\begin{equation}
F_{\Lambda }\left( X\left( z\right) \right) =\frac{\partial F\left( X\left(
z\right) \right) }{\partial X^{\Lambda }}.  \label{prepotential}
\end{equation}
In $\mathcal{N}=2$, $d=4$ supergravity the holomorphic function $F$ is
constrained to be homogeneous of degree $2$ in the contravariant symplectic
holomorphic sections $X^{\Lambda }\left( z\right) $, \textit{i.e.} (see
\textit{e.g.} \cite{3,4} and Refs. therein)
\begin{equation}
2F\left( X\left( z\right) \right) =X^{\Lambda }\left( z\right) F_{\Lambda
}\left( X\left( z\right) \right) .  \label{hom-prop-F}
\end{equation}

The normalization of the holomorphic period vector $\Pi \left( z\right) $ is
such that
\begin{eqnarray}
K\left( z,\overline{z}\right) &=&-ln\left[ i\left\langle \Pi \left( z\right)
,\overline{\Pi }\left( \overline{z}\right) \right\rangle \right] \equiv -ln%
\left[ i\Pi ^{T}\left( z\right) \Omega \overline{\Pi }\left( \overline{z}%
\right) \right] =  \notag \\
&&  \notag \\
&=&-ln\left\{ i\left[ \overline{X}^{\Lambda }\left( \overline{z}\right)
F_{\Lambda }\left( z\right) -X^{\Lambda }\left( z\right) \overline{F}%
_{\Lambda }\left( \overline{z}\right) \right] \right\} ,  \label{norm-PI}
\end{eqnarray}
where $\left\langle \cdot ,\cdot \right\rangle $ stands for the symplectic
scalar product defined by $\Omega $. Note that under a K\"{a}hler
transformation
\begin{equation}
K\left( z,\overline{z}\right) \longrightarrow K\left( z,\overline{z}\right)
+f\left( z\right) +\overline{f}\left( \overline{z}\right)
\end{equation}
($f\left( z\right) $ being a generic holomorphic function), the holomorphic
period vector transforms as
\begin{equation}
\Pi \left( z\right) \longrightarrow \Pi \left( z\right) e^{-f\left( z\right)
}\Longrightarrow X^{\Lambda }\left( z\right) \longrightarrow X^{\Lambda
}\left( z\right) e^{-f\left( z\right) .}
\end{equation}
This yields that, \textit{at least locally}, the contravariant holomorphic
symplectic sections $X^{\Lambda }\left( z\right) $ can be regarded as a set
of homogeneous coordinates on $\mathcal{M}_{n_{V}}$, provided that the
Jacobian complex $n_{V}\times n_{V}$ holomorphic matrix
\begin{equation}
e_{i}^{a}\left( z\right) \equiv \frac{\partial }{\partial z^{i}}\left( \frac{%
X^{a}\left( z\right) }{X^{0}\left( z\right) }\right) ,\text{ ~}a=1,...,n_{V}
\end{equation}
is\textit{\ invertible}. If this is the case, then one can introduce the
local projective symplectic coordinates
\begin{equation}
t^{a}\left( z\right) \equiv \frac{X^{a}\left( z\right) }{X^{0}\left(
z\right) },
\end{equation}
and the SK geometry of $\mathcal{M}_{n_{V}}$ turns out to be based on the
holomorphic prepotential $\mathcal{F}\left( t\right) \equiv \left(
X^{0}\right) ^{-2}F\left( X\right) $. By using the $t$-coordinates, Eq. (\ref
{norm-PI}) can be rewritten as follows ($\mathcal{F}_{a}\left( t\right)
=\partial _{a}\mathcal{F}\left( t\right) $, $\overline{t}^{a}=\overline{t^{a}%
}$, $\overline{\mathcal{F}}_{a}\left( \overline{t}\right) =\overline{%
\mathcal{F}_{a}\left( t\right) }$):
\begin{equation}
K\left( t,\overline{t}\right) =-ln\left\{ i\left| X^{0}\left( z\left(
t\right) \right) \right| ^{2}\left[ 2\left( \mathcal{F}\left( t\right) -%
\overline{\mathcal{F}}\left( \overline{t}\right) \right) -\left( t^{a}-%
\overline{t}^{a}\right) \left( \mathcal{F}_{a}\left( t\right) +\overline{%
\mathcal{F}}_{a}\left( \overline{t}\right) \right) \right] \right\} .
\end{equation}
By performing a K\"{a}hler gauge-fixing with $f\left( z\right) =ln\left(
X^{0}\left( z\right) \right) $, yielding that $X^{0}\left( z\right)
\longrightarrow 1$, one thus gets
\begin{equation}
\left. K\left( t,\overline{t}\right) \right| _{X^{0}\left( z\right)
\longrightarrow 1}=-ln\left\{ i\left[ 2\left( \mathcal{F}\left( t\right) -%
\overline{\mathcal{F}}\left( \overline{t}\right) \right) -\left( t^{a}-%
\overline{t}^{a}\right) \left( \mathcal{F}_{a}\left( t\right) +\overline{%
\mathcal{F}}_{a}\left( \overline{t}\right) \right) \right] \right\} .
\label{K-t-X0=1}
\end{equation}
In particular, one can choose the so-called \textit{special coordinates},
\textit{i.e.} the system of local projective $t$-coordinates such that
\begin{equation}
e_{i}^{a}\left( z\right) =\delta _{i}^{a}\Leftrightarrow t^{a}\left(
z\right) =z^{i}\left( +c^{i},\text{~}c^{i}\in \mathbb{C}\right) .
\end{equation}
Thus, Eq. (\ref{K-t-X0=1}) acquires the form
\begin{equation}
\left. K\left( t,\overline{t}\right) \right| _{X^{0}\left( z\right)
\longrightarrow 1,e_{i}^{a}\left( z\right) =\delta _{i}^{a}}=-ln\left\{ i%
\left[ 2\left( \mathcal{F}\left( z\right) -\overline{\mathcal{F}}\left(
\overline{z}\right) \right) -\left( z^{j}-\overline{z}^{\overline{j}}\right)
\left( \mathcal{F}_{j}\left( z\right) +\overline{\mathcal{F}}_{\overline{j}%
}\left( \overline{z}\right) \right) \right] \right\} .
\end{equation}

Moreover, it should be recalled that $Z$ has K\"{a}hler weights $\left( p,%
\overline{p}\right) =\left( 1,-1\right) $, and therefore its
K\"{a}hler-covariant derivatives read
\begin{equation}
D_{i}Z=\left( \partial _{i}+\frac{1}{2}\partial _{i}K\right) Z,~~\overline{D}%
_{\overline{i}}Z=\left( \overline{\partial }_{\overline{i}}-\frac{1}{2}%
\overline{\partial }_{\overline{i}}K\right) Z.  \label{DiZ}
\end{equation}

The \textit{fundamental differential relations} of SK geometry are (see
\textit{e.g.} \cite{4}; for elucidations about the various equivalent
approaches to SK geometry, see also \cite{Craps1} and \cite{Craps2}):
\begin{equation}
\left\{
\begin{array}{l}
D_{i}Z=Z_{i}~\text{(definition of \textit{matter charges})}; \\
\\
D_{i}Z_{j}=iC_{ijk}g^{k\overline{k}}\overline{D}_{\overline{k}}\overline{Z}%
=iC_{ijk}g^{k\overline{k}}\overline{Z}_{\overline{k}}; \\
\\
D_{i}\overline{D}_{\overline{j}}\overline{Z}=D_{i}\overline{Z}_{\overline{j}%
}=g_{i\overline{j}}\overline{Z}; \\
\\
D_{i}\overline{Z}=0~\text{(\textit{K\"{a}hler-covariant holomorphicity})}.
\end{array}
\right.  \label{SKG-rels1}
\end{equation}
The first relation is nothing but the definition of the so-called \textit{%
matter charges} $Z_{i}$, and the fourth relation expresses the
K\"{a}hler-covariant holomorphicity of $Z$. $C_{ijk}$ is the rank-3,
completely symmetric, covariantly holomorphic tensor of SK geometry (with
K\"{a}hler weights $\left( 2,-2\right) $) (see \textit{e.g.} \cite
{4,Castellani1,DFF}):
\begin{equation}
\begin{array}{l}
\left\{
\begin{array}{l}
C_{ijk}=\left\langle D_{i}D_{j}V,D_{k}V\right\rangle =e^{K}\left( \partial
_{i}\mathcal{N}_{\Lambda \Sigma }\right) D_{j}X^{\Lambda }D_{k}X^{\Sigma }=
\\
\\
=e^{K}\left( \partial _{i}X^{\Lambda }\right) \left( \partial _{j}X^{\Sigma
}\right) \left( \partial _{k}X^{\Xi }\right) \partial _{\Xi }\partial
_{\Sigma }F_{\Lambda }\left( X\right) \equiv e^{K}W_{ijk},\text{~~}\overline{%
\partial }_{\overline{l}}W_{ijk}=0; \\
\\
C_{ijk}=D_{i}D_{j}D_{k}\mathcal{S},~~\mathcal{S}\equiv -iL^{\Lambda
}L^{\Sigma }Im\left( F_{\Lambda \Sigma }\right) ,~~F_{\Lambda \Sigma }\equiv
\frac{\partial F_{\Lambda }}{\partial X^{\Sigma }},F_{\Lambda \Sigma }\equiv
F_{\left( \Lambda \Sigma \right) }~; \\
\\
C_{ijk}=-ig_{i\overline{l}}\overline{f}_{\Lambda }^{\overline{l}%
}D_{j}D_{k}L^{\Lambda },~~~\overline{f}_{\Lambda }^{\overline{l}}\left(
\overline{D}\overline{L}_{\overline{s}}^{\Lambda }\right) \equiv \delta _{%
\overline{s}}^{\overline{l}};
\end{array}
\right. \\
\\
\overline{D}_{\overline{i}}C_{jkl}=0\text{ (\textit{covariant holomorphicity}%
)}; \\
\\
R_{i\overline{j}k\overline{l}}=-g_{i\overline{j}}g_{k\overline{l}}-g_{i%
\overline{l}}g_{k\overline{j}}+C_{ikp}\overline{C}_{\overline{j}\overline{l}%
\overline{p}}g^{p\overline{p}}\text{ (usually named \textit{SK geometry
constraints})}; \\
\\
D_{[i}C_{j]kl}=0.
\end{array}
\label{C}
\end{equation}
the last property being a consequence, through the \textit{SK geometry
constraints} and the covariant holomorphicity of $C_{ijk}$, of the Bianchi
identities satisfied by the Riemann tensor $R_{i\overline{j}k\overline{l}}$.
As usual, square brackets denote antisymmetrization with respect to enclosed
indices.

It is worth remarking that the third of Eqs. (\ref{C}) correctly defines the
Riemann tensor $R_{i\overline{j}k\overline{l}}$, and it is actual the
opposite of the one which may be found in a large part of existing
literature. Such a formulation of the so-called \textit{SK geometry
constraints} is well defined, because it consistently yields \textit{negative%
} values of the constant scalar curvature of symmetric SK manifolds (see
\textit{e.g.} \cite{Helgason}). Furthermore, it should be recalled that in a
generic K\"{a}hler geometry $R_{i\overline{j}k\overline{l}}$ reads (see
\textit{e.g.} \cite{Zumino})
\begin{equation}
\begin{array}{l}
R_{i\overline{j}k\overline{l}}=g^{m\overline{n}}\left( \overline{\partial }_{%
\overline{l}}\overline{\partial }_{\overline{j}}\partial _{m}K\right)
\partial _{i}\overline{\partial }_{\overline{n}}\partial _{k}K-\overline{%
\partial }_{\overline{l}}\partial _{i}\overline{\partial }_{\overline{j}%
}\partial _{k}K=g_{k\overline{n}}\partial _{i}\overline{\Gamma }_{\overline{l%
}\overline{j}}^{~~\overline{n}}=g_{n\overline{l}}\overline{\partial }_{%
\overline{j}}\Gamma _{ki}^{~~n}, \\
\\
\overline{R_{i\overline{j}k\overline{l}}}=R_{j\overline{i}l\overline{k}}%
\text{ \ \ \ (\textit{reality})}, \\
\\
\Gamma _{ij}^{~~l}=-g^{l\overline{l}}\partial _{i}g_{j\overline{l}}=-g^{l%
\overline{l}}\partial _{i}\overline{\partial }_{\overline{l}}\partial
_{j}K=\Gamma _{\left( ij\right) }^{~~l},
\end{array}
\text{\ }
\end{equation}
where $\Gamma _{ij}^{~~l}$ stand for the Christoffel symbols of the second
kind of the K\"{a}hler metric $g_{i\overline{j}}$.

In the first of Eqs. (\ref{C}), a fundamental entity, the so-called kinetic
matrix $\mathcal{N}_{\Lambda \Sigma }\left( z,\overline{z}\right) $ of $%
\mathcal{N}=2$, $d=4$ supergravity, has been introduced. It is an $\left(
n_{V}+1\right) \times \left( n_{V}+1\right) $ complex symmetric,
moduli-dependent, K\"{a}hler gauge-invariant matrix defined by the following
fundamental \textit{Ans\"{a}tze}, solving the \textit{SKG constraints} given
by the third of Eqs. (\ref{C}):
\begin{equation}
M_{\Lambda }=\mathcal{N}_{\Lambda \Sigma }L^{\Sigma },~~D_{i}M_{\Lambda }=%
\overline{\mathcal{N}}_{\Lambda \Sigma }D_{i}L^{\Sigma }.  \label{Ans1}
\end{equation}
By introducing the $\left( n_{V}+1\right) \times \left( n_{V}+1\right) $
complex matrices ($I=1,...,n_{V}+1$)
\begin{equation}
f_{I}^{\Lambda }\left( z,\overline{z}\right) \equiv \left( \overline{D}_{%
\overline{i}}\overline{L}^{\Lambda }\left( z,\overline{z}\right) ,L^{\Lambda
}\left( z,\overline{z}\right) \right) ,\text{ \ }h_{I\Lambda }\left( z,%
\overline{z}\right) \equiv \left( \overline{D}_{\overline{i}}\overline{M}%
_{\Lambda }\left( z,\overline{z}\right) ,M_{\Lambda }\left( z,\overline{z}%
\right) \right) ,
\end{equation}
the \textit{Ans\"{a}tze} (\ref{Ans1}) uniquely determine $\mathcal{N}%
_{\Lambda \Sigma }\left( z,\overline{z}\right) $ as
\begin{equation}
\mathcal{N}_{\Lambda \Sigma }\left( z,\overline{z}\right) =h_{I\Lambda
}\left( z,\overline{z}\right) \circ \left( f^{-1}\right) _{\Sigma
}^{I}\left( z,\overline{z}\right) ,
\end{equation}
where $\circ $ denotes the usual matrix product, and $\left( f^{-1}\right)
_{\Sigma }^{I}f_{I}^{\Lambda }=\delta _{\Sigma }^{\Lambda }$, $\left(
f^{-1}\right) _{\Lambda }^{I}f_{J}^{\Lambda }=\delta _{J}^{I}$.

The covariantly holomorphic $\left( 2n_{V}+2\right) \times 1$ period vector $%
V\left( z,\overline{z}\right) $ is \textit{symplectically orthogonal} to all
its K\"{a}hler-covariant derivatives:
\begin{equation}
\left\{
\begin{array}{l}
\left\langle V\left( z,\overline{z}\right) ,D_{i}V\left( z,\overline{z}%
\right) \right\rangle =0; \\
\\
\left\langle V\left( z,\overline{z}\right) ,\overline{D}_{\overline{i}%
}V\left( z,\overline{z}\right) \right\rangle =0; \\
\\
\left\langle V\left( z,\overline{z}\right) ,D_{i}\overline{V}\left( z,%
\overline{z}\right) \right\rangle =0; \\
\\
\left\langle V\left( z,\overline{z}\right) ,\overline{D}_{\overline{i}}%
\overline{V}\left( z,\overline{z}\right) \right\rangle =0.
\end{array}
\right.  \label{ortho-rels}
\end{equation}
Moreover, it holds that
\begin{eqnarray}
g_{i\overline{j}}\left( z,\overline{z}\right) &=&-i\left\langle D_{i}V\left(
z,\overline{z}\right) ,\overline{D}_{\overline{j}}\overline{V}\left( z,%
\overline{z}\right) \right\rangle =  \notag \\
&=&-2Im\left( \mathcal{N}_{\Lambda \Sigma }\left( z,\overline{z}\right)
\right) D_{i}L^{\Lambda }\left( z,\overline{z}\right) \overline{D}_{%
\overline{i}}\overline{L}^{\Sigma }\left( z,\overline{z}\right) =  \notag \\
&=&2Im\left( F_{\Lambda \Sigma }\left( z\right) \right) D_{i}L^{\Lambda
}\left( z,\overline{z}\right) \overline{D}_{\overline{i}}\overline{L}%
^{\Sigma }\left( z,\overline{z}\right) ;  \label{ortho1} \\
&&  \notag \\
\left\langle V\left( z,\overline{z}\right) ,D_{i}\overline{D}_{\overline{j}%
}V\left( z,\overline{z}\right) \right\rangle &=&iC_{ijk}g^{k\overline{k}%
}\left\langle V\left( z,\overline{z}\right) ,\overline{D}_{\overline{k}}%
\overline{V}\left( z,\overline{z}\right) \right\rangle =0.  \label{ortho2}
\end{eqnarray}

The fundamental $\left( 2n_{V}+2\right) \times 1$ vector identity defining
the geometric structure of SK manifolds read as follows \cite
{FBC,K1,K2,BFM,AoB-book,K2-bis}:
\begin{equation}
Q^{T}-i\Omega \mathcal{M}\left( \mathcal{N}\right) Q^{T}=-2iZ\overline{V}%
-2ig^{j\overline{j}}\left( \overline{D}_{\overline{j}}\overline{Z}\right)
D_{j}V.  \label{SKG-identities1}
\end{equation}
The $\left( 2n_{V}+2\right) \times \left( 2n_{V}+2\right) $ real symmetric
matrix $\mathcal{M}\left( \mathcal{N}\right) $ is defined as \cite{4,FK1,FK2}
\begin{eqnarray}
\mathcal{M}\left( \mathcal{N}\right) &=&\mathcal{M}\left( Re\left( \mathcal{N%
}\right) ,Im\left( \mathcal{N}\right) \right) \equiv  \notag \\
&&  \notag \\
&\equiv &\left(
\begin{array}{ccc}
Im\left( \mathcal{N}\right) +Re\left( \mathcal{N}\right) \left( Im\left(
\mathcal{N}\right) \right) ^{-1}Re\left( \mathcal{N}\right) &  & -Re\left(
\mathcal{N}\right) \left( Im\left( \mathcal{N}\right) \right) ^{-1} \\
&  &  \\
-\left( Im\left( \mathcal{N}\right) \right) ^{-1}Re\left( \mathcal{N}\right)
&  & \left( Im\left( \mathcal{N}\right) \right) ^{-1}
\end{array}
\right) .  \notag \\
&&
\end{eqnarray}
It is worth reminding that $\mathcal{M}\left( \mathcal{N}\right) $ is
symplectic with respect to the metric $\Omega $ defined in Eq. (\ref{Omega}%
), \textit{i.e.} it satisfies ($\left( \mathcal{M}\left( \mathcal{N}\right)
\right) ^{T}=\mathcal{M}\left( \mathcal{N}\right) $)
\begin{equation}
\mathcal{M}\left( \mathcal{N}\right) \Omega \mathcal{M}\left( \mathcal{N}%
\right) =\Omega .
\end{equation}

By using Eqs. (\ref{norm-PI}), (\ref{ortho-rels}), (\ref{ortho1}) and (\ref
{ortho2}), the identity (\ref{SKG-identities1}) implies the following
relations:
\begin{equation}
\left\{
\begin{array}{l}
\left\langle V,Q^{T}-i\Omega \mathcal{M}\left( \mathcal{N}\right)
Q^{T}\right\rangle =-2Z; \\
\\
\left\langle \overline{V},Q^{T}-i\Omega \mathcal{M}\left( \mathcal{N}\right)
Q^{T}\right\rangle =0; \\
\\
\left\langle D_{i}V,Q^{T}-i\Omega \mathcal{M}\left( \mathcal{N}\right)
Q^{T}\right\rangle =0; \\
\\
\left\langle \overline{D}_{\overline{i}}\overline{V},Q^{T}-i\Omega \mathcal{M%
}\left( \mathcal{N}\right) Q^{T}\right\rangle =-2\overline{D}_{\overline{i}}%
\overline{Z}.
\end{array}
\right.  \label{SKG-SKG-yielded}
\end{equation}

There are only $2n_{V}$ independent real relations out of the $4n_{V}+4$
real ones yielded by the $2n_{V}+2$ complex identities (\ref{SKG-identities1}%
). Indeed, by taking the real and imaginary part of the vector identity (\ref
{SKG-identities1}) one respectively obtains
\begin{eqnarray}
Q^{T} &=&-2Re\left[ iZ\overline{V}+iG^{j\overline{j}}\left( \overline{D}_{%
\overline{j}}\overline{Z}\right) D_{j}V\right] =-2Im\left[ \overline{Z}V+G^{j%
\overline{j}}\left( D_{j}Z\right) \left( \overline{D}_{\overline{j}}%
\overline{V}\right) \right] ;  \notag \\
&&  \label{Re-SKG-SKG} \\
\Omega \mathcal{M}\left( \mathcal{N}\right) Q^{T} &=&2Im\left[ iZ\overline{V}%
+iG^{j\overline{j}}\left( \overline{D}_{\overline{j}}\overline{Z}\right)
D_{j}V\right] =2Re\left[ \overline{Z}V+G^{j\overline{j}}\left( D_{j}Z\right)
\left( \overline{D}_{\overline{j}}\overline{V}\right) \right] .  \notag \\
&&  \label{Im-SKG-SKG}
\end{eqnarray}
Consequently, the imaginary and real parts of the vector identity (\ref
{SKG-identities1}) are \textit{linearly dependent} one from the other, being
related by the $\left( 2n_{V}+2\right) \times \left( 2n_{V}+2\right) $ real
matrix
\begin{equation}
\Omega \mathcal{M}\left( \mathcal{N}\right) =\left(
\begin{array}{ccc}
\left( Im\left( \mathcal{N}\right) \right) ^{-1}Re\left( \mathcal{N}\right)
&  & -\left( Im\left( \mathcal{N}\right) \right) ^{-1} \\
&  &  \\
Im\left( \mathcal{N}\right) +Re\left( \mathcal{N}\right) \left( Im\left(
\mathcal{N}\right) \right) ^{-1}Re\left( \mathcal{N}\right) &  & -Re\left(
\mathcal{N}\right) \left( Im\left( \mathcal{N}\right) \right) ^{-1}
\end{array}
\right) .  \label{epsilon-emme}
\end{equation}
Put another way, Eqs. (\ref{Re-SKG-SKG}) and (\ref{Im-SKG-SKG}) yield
\begin{equation}
Re\left[ Z\overline{V}+G^{j\overline{j}}\left( \overline{D}_{\overline{j}}%
\overline{Z}\right) D_{j}V\right] =\Omega \mathcal{M}\left( \mathcal{N}%
\right) Im\left[ Z\overline{V}+G^{j\overline{j}}\left( \overline{D}_{%
\overline{j}}\overline{Z}\right) D_{j}V\right] ,  \label{rotation1}
\end{equation}
expressing the fact that the real and imaginary parts of the quantity $Z%
\overline{V}+G^{j\overline{j}}\left( \overline{D}_{\overline{j}}\overline{Z}%
\right) D_{j}V$ are simply related through a \textit{symplectic rotation}
given by the matrix $\Omega \mathcal{M}\left( \mathcal{N}\right) $, whose
simplecticity directly follows from the symplectic nature of $\mathcal{M}%
\left( \mathcal{N}\right) $. Eq. (\ref{rotation1}) reduces the number of
independent real relations implied by the identity (\ref{SKG-identities1})
from $4n_{V}+4$ to $2n_{V}+2$.

Moreover, it should be stressed that vector identity (\ref{SKG-identities1})
entails $2$ \textit{redundant} degrees of freedom, encoded in the
homogeneity (of degree $1$) of (\ref{SKG-identities1}) under \textit{complex
scalings} of $Q$. Indeed, by using the definition (\ref{Z}), it is easy to
check that the right-hand side of (\ref{SKG-identities1}) gets scaled by an
overall factor $\lambda $ under the following transformation on $Q$:
\begin{equation}
Q\longrightarrow \lambda Q,~~~\lambda \in \mathbb{C}.
\end{equation}
Thus, as announced, only $2n_{V}$ real independent relations are actually
yielded by the vector identity (\ref{SKG-identities1}).

This is clearly consistent with the fact that the $2n_{V}+2$\ complex
identities (\ref{SKG-identities1}) express nothing but a \textit{change of
basis} of the BH charge configurations, between the K\"{a}hler-invariant $%
1\times \left( 2n_{V}+2\right) $\ symplectic (magnetic/electric) basis
vector $Q$ defined by Eq. (\ref{Gamma-tilde}) and the complex,
moduli-dependent $1\times \left( n_{V}+1\right) $ \textit{matter charges}
vector (with K\"{a}hler weights $\left( 1,-1\right) $)
\begin{equation}
\mathcal{Z}\left( z,\overline{z}\right) \equiv \left( Z\left( z,\overline{z}%
\right) ,Z_{i}\left( z,\overline{z}\right) \right) _{i=1,...,n_{V}}.
\label{Z-call}
\end{equation}

It should be recalled that the BH charges are conserved due to the overall $%
\left( U(1)\right) ^{n_{V}+1}$ gauge-invariance of the system under
consideration, and $Q$ and $\mathcal{Z}\left( z,\overline{z}\right) $ are
two \textit{equivalent} basis for them. Their very equivalence relations are
given by the identities (\ref{SKG-identities1}) themselves. By its very
definition (\ref{Gamma-tilde}), $Q$\ is \textit{moduli-independent} (at
least in a static, spherically symmetric and asymptotically flat extremal BH
background, as it is the case being treated here), whereas $Z$ is \textit{%
moduli-dependent}, since it refers to the \textit{eigenstates} of the $%
\mathcal{N}=2$, $d=4$ supergravity multiplet and of the $n_{V}$\ Maxwell
vector multiplets.

\subsection{\label{Sect2}$\frac{1}{2}$-BPS Attractors}

In $\mathcal{N}=2$, $d=4$ supergravity the following expression holds \cite
{FK1,FK2,4}:
\begin{equation}
V_{BH}=\left| Z\right| ^{2}+g^{j\overline{j}}\left( D_{j}Z\right) \overline{D%
}_{\overline{j}}\overline{Z}.  \label{VBH1}
\end{equation}
An elegant way to obtain $V_{BH}$ is given by left-multiplying the vector
identity (\ref{SKG-identities1}) by the $1\times \left( 2n_{V}+2\right) $
complex moduli-dependent vector $-\frac{1}{2}Q\mathcal{M}\left( \mathcal{N}%
\right) $; due to the symplecticity of the matrix $\mathcal{M}\left(
\mathcal{N}\right) $, one obtains \cite{FK1,FK2,4}
\begin{equation}
V_{BH}=-\frac{1}{2}Q\mathcal{M}\left( \mathcal{N}\right) Q^{T}.  \label{VBH2}
\end{equation}
Thus, $V_{BH}$ is identified with the first (of two), lowest-order (-
\textit{quadratic }- in charges), \textit{positive-definite} real invariant $%
I_{1}$ of SK geometry (see \textit{e.g.} \cite{K2-bis,4}).\ It is worth
noticing that the result (\ref{VBH2}) can also be derived from the SK
geometry identities (\ref{SKG-identities1}) by using the relation (see \cite
{FKlast}, where a generalization for $\mathcal{N}>2$-extended supergravities
is also given)\textbf{\ }
\begin{equation}
\frac{1}{2}\left( \mathcal{M}\left( \mathcal{N}\right) +i\Omega \right)
\mathcal{V}=i\Omega \mathcal{V}\Leftrightarrow \mathcal{M}\left( \mathcal{N}%
\right) \mathcal{V}=i\Omega \mathcal{V},
\end{equation}
where $\mathcal{V}$ is a $\left( 2n_{V}+2\right) \times \left(
n_{V}+1\right) $ matrix defined as:
\begin{equation}
\mathcal{V}\equiv \left( V,\overline{D}_{\overline{1}}\overline{V},...,%
\overline{D}_{\overline{n_{V}}}\overline{V}\right) .
\end{equation}

By differentiating Eq. (\ref{VBH1}) with respect to the scalars, it is easy
to check that the general criticality conditions (\ref{crit-cond-gen}) can
be recast in the following form \cite{FGK}:
\begin{equation}
D_{i}V_{BH}=\partial _{i}V_{BH}=0\Leftrightarrow 2\overline{Z}D_{i}Z+g^{j%
\overline{j}}\left( D_{i}D_{j}Z\right) \overline{D}_{\overline{j}}\overline{Z%
}=0;  \label{AEs1}
\end{equation}
this is what one should rigorously call the $\mathcal{N}=2$, $d=4$ \textit{%
Attractor Eqs.} (AEs). By means of the features of SK geometry given by Eqs.
(\ref{SKG-rels1}), the $\mathcal{N}=2$ AEs (\ref{AEs1}) can be re-expressed
as follows \cite{FGK}:
\begin{equation}
2\overline{Z}Z_{i}+iC_{ijk}g^{j\overline{j}}g^{k\overline{k}}\overline{Z}_{%
\overline{j}}\overline{Z}_{\overline{k}}=0.  \label{AEs2}
\end{equation}
It is evident that the tensor $C_{ijk}$ is crucial in relating the $\mathcal{%
N}=2$ central charge function $Z$ (\textit{graviphoton charge}) and the $%
n_{V}$ \textit{matter charges} $Z_{i}$ (coming from the $n_{V}$ Abelian
vector multiplets) at the critical points of $V_{BH}$ in the SK scalar
manifold $\mathcal{M}_{n_{V}}$.

The static, spherically symmetric, asymptotically flat dyonic (not
necessarily extremal) $d=4$ BHs are known to be described by an \textit{%
effective} $d=1$ Lagrangian (\cite{FGK}, \cite{2}, and also \cite{AoB-book}
and \cite{ADFT}), with $V_{BH}$ and effective fermionic \textit{``mass
terms''} controlled by the vector $Q$ defined by Eq. (\ref{Gamma-tilde}).
The \textit{``apparent'' gravitino mass} is given by $Z$, whereas the
\textit{gaugino mass matrix} $\Lambda _{ij}$ reads (see the second Ref. of
\cite{DFF})
\begin{equation}
\Lambda _{ij}=C_{ijk}g^{k\overline{k}}\overline{Z}_{\overline{k}}.
\end{equation}
The \textit{supersymmetry breaking order parameters}, related to the \textit{%
mixed gravitino-gaugino couplings}, are nothing but the \textit{matter
charge( function)s} $D_{i}Z=Z_{i}$ (see the first of Eqs. (\ref{SKG-rels1})).

As evident from the AEs (\ref{AEs1}) and (\ref{AEs2}), the conditions
\begin{equation}
(Z\neq 0,)~D_{i}Z=0\text{~~~}\forall i=1,...,n_{V}  \label{BPS-conds}
\end{equation}
determine a (\textit{non-degenerate}) critical point of $V_{BH}$, namely a $%
\frac{1}{2}$-BPS critical point, which preserve four supersymmetry degrees
of freedom out of the eight pertaining to the $\mathcal{N}=2$, $d=4$
Poincar\'{e} superalgebra related to the asymptotical Minkowski background.
The corresponding Bekenstein entropy reads \cite{FKS}-\nocite{Strom,FK1,FK2}
\cite{FGK}:\textbf{\ }
\begin{equation}
S_{BH,\frac{1}{2}-BPS}=\pi \left. V_{BH}\right| _{\frac{1}{2}-BPS}=\pi
\left\{ \left| Z\right| _{\frac{1}{2}-BPS}^{2}+\left[ g^{i\overline{i}%
}\left( D_{i}Z\right) \left( \overline{D}_{\overline{i}}\overline{Z}\right) %
\right] _{\frac{1}{2}-BPS}\right\} =\left| Z\right| _{\frac{1}{2}-BPS}^{2}>0.
\label{V-BPS}
\end{equation}

In general, $\frac{1}{2}$-BPS critical points are (\textit{at least local})
\textit{minima} of $V_{BH}$ in $\mathcal{M}_{n_{V}}$, and therefore they are
\textit{stable}; thus, they are attractors in a strict sense. Indeed, the $%
2n_{V}\times 2n_{V}$ matrix $\mathcal{H}^{V_{BH}}$ (within the K\"{a}hler
holomorphic/antiholomorphic parametrization) evaluated at such points is
\textit{strictly positive-definite} \cite{FGK}:
\begin{eqnarray}
&&
\begin{array}{l}
\left( D_{i}D_{j}V_{BH}\right) _{\frac{1}{2}-BPS}=\left( \partial
_{i}\partial _{j}V_{BH}\right) _{\frac{1}{2}-BPS}=0, \\
\\
\left( D_{i}\overline{D}_{\overline{j}}V_{BH}\right) _{\frac{1}{2}%
-BPS}=\left( \partial _{i}\overline{\partial }_{\overline{j}}V_{BH}\right) _{%
\frac{1}{2}-BPS}=2\left( g_{i\overline{j}}V_{BH}\right) _{\frac{1}{2}%
-BPS}=2\left. g_{i\overline{j}}\right| _{\frac{1}{2}-BPS}\left| Z\right| _{%
\frac{1}{2}-BPS}^{2}>0,
\end{array}
\notag \\
&&  \label{SUSY-crit}
\end{eqnarray}
wherethe notation ``$>0$'' is here understood as strict
positive-definiteness. Eqs. (\ref{SUSY-crit}) yield that the Hermiticity and
(strict) positive-definiteness of $\mathcal{H}^{V_{BH}}$ (in $\left( z,%
\overline{z}\right) $-coordinates) at the $\frac{1}{2}$-BPS critical points
are due to the Hermiticity and - assumed - (strict) positive-definiteness
(actually holding \textit{globally}) of the metric $g_{i\overline{j}}$ of $%
\mathcal{M}_{n_{V}}$.

Considering the $\mathcal{N}=2$, $d=4$ supergravity Lagrangian in a static,
spherically symmetric, asymptotically flat dyonic BH background, and
denoting by $\psi $ and $\lambda ^{i}$ respectively the gravitino and
gaugino fields, it is easy to see that such a Lagrangian contains terms of
the form (see the second and third Refs. of \cite{DFF})
\begin{equation}
\begin{array}{l}
Z\psi \psi ; \\
\\
C_{ijk}g^{k\overline{k}}\left( \overline{D}_{\overline{k}}\overline{Z}%
\right) \lambda ^{i}\lambda ^{j}; \\
\\
\left( D_{i}Z\right) \lambda ^{i}\psi .
\end{array}
\end{equation}
Thus, the $\frac{1}{2}$-BPS conditions (\ref{BPS-conds}) implies the gaugino
mass term and the $\lambda \psi $ term to vanish at the $\frac{1}{2}$-BPS
critical points of $V_{BH}$ in $\mathcal{M}_{n_{V}}$. It is interesting to
remark that the gravitino \textit{``apparent mass''} term $Z\psi \psi $ is
in general \textit{non-vanishing}, also when evaluated at the considered $%
\frac{1}{2}$-BPS attractors; this is ultimately a consequence of the fact
that the extremal BH horizon geometry at the $\frac{1}{2}$-BPS (as well as
at the non-BPS) attractors is \textit{Bertotti-Robinson} $AdS_{2}\times
S^{2} $ (with vanishing scalar curvature and \textit{conformally flat}) \cite
{BR1,BR2,BR3}.

\subsection{\label{Sect3}Non-BPS $Z\neq 0$ Critical Points of $V_{BH}$}

The $\frac{1}{2}$-BPS conditions (\ref{BPS-conds}) are \textit{not} the most
general ones solving the $\mathcal{N}=2$, $d=4$ AEs (\ref{AEs1}) or (\ref
{AEs2}). For instance, one might consider critical points of $V_{BH}$ (thus
satisfying the AEs (\ref{AEs1}) or (\ref{AEs2})) characterized by
\begin{equation}
\left\{
\begin{array}{l}
D_{i}Z\neq 0,\text{ for \textit{(at least one)} }i, \\
\\
Z\neq 0.
\end{array}
\right.  \label{non-BPS-Z<>0}
\end{equation}
Such critical points are \textit{non-supersymmetric} ones (\textit{i.e.}
they do \textit{not} preserve any of the eight supersymmetry degrees of
freedom of the asymptotical Minkowski background), and they correspond to an
extremal,\textit{\ non-BPS} BH background. They are commonly named \textit{%
non-BPS} $Z\neq 0$ \textit{critical points of }$V_{BH}$. We will devote the
present Section to present their main features.

The corresponding non-BPS $Z\neq 0$ Bekenstein-Hawking entropy reads (\cite
{K1}, \cite{K2}, \cite{Tom}):
\begin{equation}
\begin{array}{l}
S_{BH,non-BPS,Z\neq 0}=\pi \left. V_{BH}\right| _{non-BPS,Z\neq 0}= \\
\\
=\pi \left[ \left| Z\right| _{non-BPS,Z\neq 0}^{2}+\left[ g^{i\overline{i}%
}\left( D_{i}Z\right) \left( \overline{D}_{\overline{i}}\overline{Z}\right)
\right] _{non-BPS,Z\neq 0}\right] >\pi \left| Z\right| _{non-BPS,Z\neq
0}^{2},
\end{array}
\label{V-non-BPS}
\end{equation}
not saturating the BPS bound. As implied by AEs (\ref{AEs2}), if at non-BPS $%
Z\neq 0$ critical points it holds that $D_{i}Z\neq 0$ for \textit{at least
one} index $i$ and $Z\neq 0$, then
\begin{equation}
\left( C_{ijk}\right) _{non-BPS,Z\neq 0}\neq 0,~~~\text{\textit{for some} }%
\left( i,j,k\right) \in \left\{ 1,...,n_{V}\right\} ^{3},
\end{equation}
\textit{i.e.} the rank-3 symmetric tensor $C_{ijk}$ will for sure have some
non-vanishing components in order for criticality conditions (\ref{AEs2})\
to be satisfied at non-BPS $Z\neq 0$ critical points.

Moreover, the general criticality conditions (\ref{AEs1}) for $V_{BH}$ can
be recognized to be the general \textit{Ward identities} relating the
gravitino mass $Z$, the gaugino masses $D_{i}D_{j}Z$ and the
supersymmetry-breaking order parameters $D_{i}Z$ in a generic \textit{%
spontaneously broken} supergravity theory \cite{9}. Indeed, away from $\frac{%
1}{2}$-BPS critical points (\textit{i.e.} for $D_{i}Z\neq 0$ for some $i$),
the AEs (\ref{AEs1}) can be re-expressed as follows (see also \cite
{Saraikin-Vafa-1}):
\begin{equation}
\left( \mathbf{M}_{ij}h^{j}\right) _{\partial V_{BH}=0}=0,  \label{EQ}
\end{equation}
with
\begin{equation}
\mathbf{M}_{ij}\equiv D_{i}D_{j}Z+2\frac{\overline{Z}}{\left[ g^{k\overline{k%
}}\left( D_{k}Z\right) \overline{D}_{\overline{k}}\overline{Z}\right] }%
\left( D_{i}Z\right) D_{j}Z,\text{ }(\text{\textit{K\"{a}hler weights} }%
\left( 1,-1\right) ),
\end{equation}
and
\begin{equation}
h^{j}\equiv g^{j\overline{j}}\overline{D}_{\overline{j}}\overline{Z},\text{ }%
(\text{\textit{K\"{a}hler weights} }\left( -1,1\right) ).
\end{equation}
For a non-vanishing contravariant vector $h^{j}$ (\textit{i.e. }away from $%
\frac{1}{2}$-BPS critical points, as pointed out above), Eq. (\ref{EQ})
admits a solution \textit{iff} the $n_{V}\times n_{V}$ complex symmetric
matrix $\mathbf{M}_{ij}$ has vanishing determinant (implying that it has
\textit{at most} $n_{V}-1$ non-vanishing eigenvalues) at the considered
(non-BPS) critical points of $V_{BH}$ (however, notice that $\mathbf{M}_{ij}$%
\ is symmetric but not necessarily Hermitian, thus in general its
eigenvalues are not necessarily real).

By using the properties of SK geometry, the non-BPS $Z\neq 0$
Bekenstein-Hawking entropy (\ref{V-non-BPS}) can be further elaborated as
follows \cite{review-Kallosh}:
\begin{eqnarray}
\frac{S_{BH,non-BPS,Z\neq 0}}{\pi } &=&\left\{ \left| Z\right| ^{2}\cdot %
\left[ 1+\frac{1}{4\left| Z\right| ^{4}}R_{k\overline{r}n\overline{s}}g^{k%
\overline{m}}g^{t\overline{r}}g^{n\overline{l}}g^{u\overline{s}}\left(
D_{t}Z\right) \left( D_{u}Z\right) \left( \overline{D}_{\overline{l}}%
\overline{Z}\right) \overline{D}_{\overline{m}}\overline{Z}+\right. \right.
\notag \\
&&\left. \left. +\frac{1}{2\left| Z\right| ^{4}}\left[ g^{i\overline{j}%
}\left( D_{i}Z\right) \overline{D}_{\overline{j}}\overline{Z}\right] ^{2}%
\right] \right\} _{non-BPS,Z\neq 0}.  \label{V-non-BPS-Z<>0-1}
\end{eqnarray}
One can then introduce the so-called \textit{non-BPS }$Z\neq 0$\textit{\
supersymmetry breaking order parameter} \cite{review-Kallosh}:
\begin{eqnarray}
\left( 0<\right) \mathcal{O}_{non-BPS,Z\neq 0} &\equiv &\left[ \frac{g^{i%
\overline{j}}\left( D_{i}Z\right) \overline{D}_{\overline{j}}\overline{Z}}{%
\left| Z\right| ^{2}}\right] _{non-BPS,Z\neq 0}=  \notag \\
&&  \notag \\
&=&-\left[ \frac{i}{2\overline{Z}\left| Z\right| ^{2}}C_{ijk}g^{i\overline{i}%
}g^{j\overline{l}}g^{k\overline{m}}\left( \overline{D}_{\overline{i}}%
\overline{Z}\right) \left( \overline{D}_{\overline{l}}\overline{Z}\right)
\overline{D}_{\overline{m}}\overline{Z}\right] _{non-BPS,Z\neq 0}=  \notag \\
&&  \notag \\
&=&\left[ \frac{1}{4\left| Z\right| ^{4}}g^{i\overline{j}}C_{ikn}\overline{C}%
_{\overline{j}\overline{r}\overline{s}}g^{n\overline{l}}g^{k\overline{m}}g^{t%
\overline{r}}g^{u\overline{s}}\left( D_{t}Z\right) \left( D_{u}Z\right)
\left( \overline{D}_{\overline{l}}\overline{Z}\right) \overline{D}_{%
\overline{m}}\overline{Z}\right] _{non-BPS,Z\neq 0}.  \notag \\
&&  \label{UCLApre}
\end{eqnarray}
Consequently
\begin{eqnarray}
S_{BH,non-BPS,Z\neq 0} &=&\pi \left\{ \left| Z\right| _{non-BPS,Z\neq 0}^{2}
\left[ 1+\mathcal{O}_{non-BPS,Z\neq 0}\right] \right\} =  \notag \\
&=&\pi \left| Z\right| _{non-BPS,Z\neq 0}^{2}\cdot  \notag \\
&&\cdot \left[ 3-2\frac{\mathcal{R}\left( Z\right) }{g^{i\overline{j}}C_{ikn}%
\overline{C}_{\overline{j}\overline{r}\overline{s}}g^{n\overline{l}}g^{k%
\overline{m}}g^{t\overline{r}}g^{u\overline{s}}\left( D_{t}Z\right) \left(
D_{u}Z\right) \left( \overline{D}_{\overline{l}}\overline{Z}\right)
\overline{D}_{\overline{m}}\overline{Z}}\right] _{non-BPS,Z\neq 0},  \notag
\\
&&  \label{CERN-lonely1}
\end{eqnarray}
where the \textit{sectional curvature} (see \textit{e.g.} \cite
{Riemann-Finsler} and \cite{Differential-Geometry})
\begin{equation}
\mathcal{R}\left( Z\right) \equiv R_{i\overline{j}k\overline{l}}g^{i%
\overline{i}}g^{j\overline{j}}g^{k\overline{k}}g^{l\overline{l}}\left(
D_{j}Z\right) \left( D_{l}Z\right) \left( \overline{D}_{\overline{i}}%
\overline{Z}\right) \overline{D}_{\overline{k}}\overline{Z}
\label{sectional-curvature}
\end{equation}
was introduced.

Now, by using the relations of SK geometry it is possible to show that \cite
{ADFT}
\begin{gather}
\overline{D}_{\overline{m}}D_{i}C_{jkl}=\left[ \overline{D}_{\overline{m}%
},D_{i}\right] C_{jkl}=\overline{D}_{\overline{m}}D_{(i}C_{j)kl}=\overline{D}%
_{\overline{m}}D_{(i}C_{jkl)}=3C_{p(kl}C_{ij)n}g^{n\overline{n}}g^{p%
\overline{p}}\overline{C}_{\overline{n}\overline{p}\overline{m}}-4g_{\left(
l\right| \overline{m}}C_{\left| ijk\right) };  \notag \\
\Updownarrow   \notag \\
C_{p(kl}C_{ij)n}g^{n\overline{n}}g^{p\overline{p}}\overline{C}_{\overline{n}%
\overline{p}\overline{m}}=\frac{4}{3}g_{\left( l\right| \overline{m}%
}C_{\left| ijk\right) }+\overline{E}_{\overline{m}\left( ijkl\right) },
\label{UCLA1}
\end{gather}
where the rank-5 \textit{E-tensor} \cite{ADFT}
\begin{gather}
\overline{E}_{\overline{m}ijkl}=\overline{E}_{\overline{m}\left( ijkl\right)
}\equiv \frac{1}{3}\overline{D}_{\overline{m}}D_{i}C_{jkl}=\frac{1}{3}%
\overline{D}_{\overline{m}}D_{(i}C_{jkl)}=C_{p(kl}C_{ij)n}g^{n\overline{n}%
}g^{p\overline{p}}\overline{C}_{\overline{n}\overline{p}\overline{m}}-\frac{4%
}{3}g_{\left( l\right| \overline{m}}C_{\left| ijk\right) }  \notag
\label{E-tensor-def} \\
=g^{n\overline{n}}R_{\left( i\right| \overline{m}\left| j\right| \overline{n}%
}C_{n\left| kl\right) }+\frac{2}{3}g_{\left( i\right| \overline{m}}C_{\left|
jkl\right) }  \notag \\
\end{gather}
was introduced. It can be shown that \cite{ADFT,review-Kallosh}
\begin{gather}
\frac{S_{BH,non-BPS,Z\neq 0}}{\pi }=\left| Z\right| _{non-BPS,Z\neq
0}^{2}\cdot   \notag \\
\notag \\
\cdot \left\{ 4-\frac{3}{4}\left[ \frac{1}{\left| Z\right| ^{2}}\frac{%
E_{i\left( \overline{k}\overline{l}\overline{m}\overline{n}\right) }g^{i%
\overline{j}}g^{k\overline{k}}g^{l\overline{l}}g^{m\overline{m}}g^{n%
\overline{n}}\left( \overline{D}_{\overline{j}}\overline{Z}\right) \left(
D_{k}Z\right) \left( D_{l}Z\right) \left( D_{m}Z\right) D_{n}Z}{N_{3}\left(
Z\right) }\right] _{non-BPS,Z\neq 0}\right\} ,  \notag \\
\end{gather}
where the \textit{complex cubic form}
\begin{equation}
N_{3}\left( Z\right) \equiv \overline{C}_{\overline{i}\overline{j}\overline{k%
}}g^{i\overline{i}}g^{j\overline{j}}g^{k\overline{k}}\left( D_{i}Z\right)
\left( D_{j}Z\right) D_{k}Z  \label{UCLA9}
\end{equation}
was introduced.

Let us now consider the case of \textit{symmetric} SK manifolds, in which
the K\"{a}hler-invariant Riemann-Christoffel tensor $R_{i\overline{j}k%
\overline{l}}$ is \textit{covariantly constant}\footnote{%
Indeed, due to the \textit{reality} of $R_{i\overline{j}k\overline{l}}$ in
any K\"{a}hler manifold, it holds that
\begin{equation*}
D_{m}R_{i\overline{j}k\overline{l}}=0\Leftrightarrow \overline{D}_{\overline{%
m}}R_{i\overline{j}k\overline{l}}=0.
\end{equation*}
}. From this it follows that \cite{CVP}: \textbf{\ }
\begin{equation}
D_{m}R_{i\overline{j}k\overline{l}}=0\Leftrightarrow
D_{i}C_{jkl}=D_{(i}C_{j)kl}=0.  \label{UCLA12}
\end{equation}
This is a \textit{sufficient} (\textit{but generally not necessary}\footnote{%
Indeed, some\textit{\ non-symmetric} SK (\textit{a priori} not necessarily
homogeneous) manifolds might exist such that $D_{(i}C_{j)kl}\neq 0$, but
however (\textit{globally}) satisfying
\begin{equation*}
\overline{D}_{\overline{m}}D_{i}C_{jkl}=\overline{D}_{\overline{m}%
}D_{(i}C_{jkl)}=\overline{\partial }_{\overline{m}}D_{(i}C_{jkl)}-\left(
\overline{\partial }_{\overline{m}}K\right) D_{(i}C_{jkl)}=0.
\end{equation*}
}) condition for the \textit{global} vanishing of the (complex conjugate)
\textit{E-tensor} $\overline{E}_{\overline{i}jklm}$:
\begin{equation}
D_{(i}C_{j)kl}=0\Rightarrow \overline{D}_{\overline{m}}D_{i}C_{jkl}=0%
\Leftrightarrow D_{m}\overline{D}_{\overline{i}}\overline{C}_{\overline{j}%
\overline{k}\overline{l}}=0,
\end{equation}
yielding \cite{CVP}
\begin{equation}
C_{p(kl}C_{ij)n}g^{n\overline{n}}g^{p\overline{p}}\overline{C}_{\overline{n}%
\overline{p}\overline{m}}=\frac{4}{3}g_{\left( l\right| \overline{m}%
}C_{\left| ijk\right) }\Leftrightarrow g^{n\overline{n}}R_{\left( i\right|
\overline{m}\left| j\right| \overline{n}}C_{n\left| kl\right) }=-\frac{2}{3}%
g_{\left( i\right| \overline{m}}C_{\left| jkl\right) }.  \label{UCLA13}
\end{equation}
Furthermore, the following noteworthy relation, holding in \textit{symmetric
SK manifolds}, can be proved \cite{review-Kallosh}:
\begin{gather}
\left( Z\left| Z\right| ^{2}\right) _{non-BPS,Z\neq 0}=\frac{i}{6}\left[
N_{3}\left( Z\right) \right] _{non-BPS,Z\neq 0}  \label{UCLA14} \\
\Downarrow  \notag \\
Re\left( \left[ \frac{N_{3}\left( Z\right) }{Z}\right] _{non-BPS,Z\neq
0}\right) =0;~Im\left( \left[ \frac{N_{3}\left( Z\right) }{Z}\right]
_{non-BPS,Z\neq 0}\right) =-6\left| Z\right| _{non-BPS,Z\neq 0}^{2}.
\end{gather}

Consequently, the \textit{supersymmetry breaking order parameter} at
non-BPS, $Z\neq 0$ critical points of $V_{BH}$ in \textit{\ symmetric} SK
manifolds is
\begin{equation}
\mathcal{O}_{non-BPS,Z\neq 0}=3,  \label{UCLA15}
\end{equation}
which might be called the \textit{``Rule of Three''} in $\mathcal{N}=2$, $%
d=4 $ supergravity (an analogous \textit{``Rule of Eight'' }seemingly exists
for symmetric \textit{real special} geometry in $d=5$ \cite{FG2}) . By
substituting into Eq. (\ref{CERN-lonely1}), one thus finally gets that
\begin{eqnarray}
\frac{S_{BH,non-BPS,Z\neq 0}}{\pi } &=&V_{BH,non-BPS,Z\neq 0}=4\left|
Z\right| _{non-BPS,Z\neq 0}^{2}=  \label{UCLA16} \\
&=&\frac{2}{3}i\left[ \frac{N_{3}\left( Z\right) }{Z}\right] _{non-BPS,Z\neq
0},  \label{UCLA16-CERN}
\end{eqnarray}
The result given by Eq. (\ref{UCLA16}) has been firstly obtained, by
exploiting group-theoretical methods, in \cite{bellucci1} (in the $stu$
model, this was firstly proved by using $\mathcal{N}=8$ arguments, in \cite
{FKlast}). Eq. (\ref{UCLA16}) also yields the following expression of the
non-BPS $Z\neq 0$ \textit{sectional curvature} of \textit{symmetric} SK
manifolds:
\begin{equation}
\left. \mathcal{R}\left( Z\right) \right| _{non-BPS,Z\neq 0}=-6\left|
Z\right| _{non-BPS,Z\neq 0}^{4}<0.  \label{UCLA17}
\end{equation}

It is worth pointing out that, while Eq. (\ref{UCLA12}) (holding\textit{\
globally}) is peculiar to \textit{symmetric} SK manifolds, Eqs. (\ref{UCLA14}%
)-(\ref{UCLA17}) actually should hold in general also for homogeneous
\textit{non-symmetric} SK manifolds, in which the Riemann-Christoffel tensor
$R_{i\overline{j}k\overline{l}}$ (and thus, through the SK constraints, $%
C_{ijk}$) is \textit{not} covariantly constant. Indeed, as obtained in \cite
{DFT07-1} \textit{at least} for all the non-BPS, $Z\neq 0$ critical points
of $V_{BH}$ considered therein, in homogeneous \textit{non-symmetric} SK
manifolds it holds that
\begin{equation}
\left[ E_{i\left( \overline{k}\overline{l}\overline{m}\overline{n}\right)
}g^{i\overline{j}}g^{k\overline{k}}g^{l\overline{l}}g^{m\overline{m}}g^{n%
\overline{n}}\left( \overline{D}_{\overline{j}}\overline{Z}\right) \left(
D_{k}Z\right) \left( D_{l}Z\right) \left( D_{m}Z\right) D_{n}Z\right]
_{non-BPS,Z\neq 0}=0,  \label{UCLA20}
\end{equation}
which seems to be the most general (necessary and sufficient) condition in
order for Eqs. (\ref{UCLA14})-(\ref{UCLA17}) to hold.

Moreover, it is worth remarking that in \cite{TT} the \textit{``Rule of
Three''} (\ref{UCLA15}) and thus
\begin{equation}
V_{BH,non-BPS,Z\neq 0}=4\left| Z\right| _{non-BPS,Z\neq 0}^{2}
\end{equation}
was proved to hold for a \textit{generic} $d$-SK geometry \cite{dWVVP},
\textit{i.e.} for a general SK geometry with a \textit{cubic} holomorphic
prepotential (for instance corresponding to the \textit{large volume limit}
of Type IIA superstrings on Calabi-Yau threefolds), for the non-BPS, $Z\neq
0 $ critical points $z_{non-BPS,Z\neq 0}^{i}$ of $V_{BH}$ supported by the
BH charge configuration with $q_{i}=0$ $\forall i$ (the one given by $%
D0-D4-D6$ brane charges in Calabi-Yau compactifications) and satisfying the
\textit{Ansatz }\cite{TT}
\begin{equation}
z_{non-BPS,Z\neq 0}^{i}=p^{i}\tau ,~\forall i=1,...,n_{V},  \label{UCLA21}
\end{equation}
where $\tau $ is quantity dependent only from the supporting BH charge
configuration.

Non-BPS $Z\neq 0$ critical points of $V_{BH}$ in $\mathcal{M}_{n_{V}}$ are
generally \textit{not necessarily stable}, because the $2n_{V}\times 2n_{V}$
matrix $\mathcal{H}^{V_{BH}}$ (within the K\"{a}hler
holomorphic/antiholomorphic parametrization) evaluated at such points is
\textit{not necessarily strictly positive-definite}. An explicit condition
of stability of non-BPS $Z\neq 0$ critical points of $V_{BH}$ has been
worked out in the $n_{V}=1$ case (see \cite{BFM}, \cite{AoB-book}, \cite
{BFMY}).

In general, the conditions (\ref{non-BPS-Z<>0}) imply the gaugino mass term,
the $\lambda \psi $ term and the gravitino \textit{``apparent mass''} term $%
Z\psi \psi $ to be \textit{non-vanishing}, when evaluated at the considered
non-BPS $Z\neq 0$ critical points of $V_{BH}$.

\section{\label{Conclusion}\textbf{Charge Orbits and Moduli Spaces of
Attractors\newline
in }$\mathcal{N}\geqslant 2$, $d=4$\textbf{\ (Symmetric) Supergravity}}

\subsection{\label{N=2-d=4-symm}$\mathcal{N}=2$, $d=4$ Symmetric Supergravity%
}

In \cite{bellucci1} the general solutions to the AEs were obtained and
classified by group-theoretical methods for those $\mathcal{N}=2$, $d=4$
supergravities having an \textit{symmetric} SK scalar manifold, \textit{i.e.}
such that $\mathcal{M}_{n_{V}}=\frac{G}{H}$, with a \textit{globally
covariantly constant }Riemann tensor $R_{i\overline{j}k\overline{l}} $: $%
D_{m}R_{i\overline{j}k\overline{l}}=0$. Such a conditions can be transported
on $C_{ijk}$ by means of the so-called \textit{SK geometry constraints} (see
the third of Eqs. (\ref{C})), obtaining $D_{l}C_{ijk}=D_{(l}C_{i)jk}=0$
(where the last of Eqs. (\ref{C}) was used).

Such $\mathcal{N}=2$, $d=4$ theories are usually named \textit{\ symmetric}
supergravities, and they have been classified in literature \cite
{CKV,GST1,GST2,GST3,GST4,CVP,dWVVP}.

With the exception of the ones based on\footnote{%
The quadratic irreducible rank-$1$ infinite sequence $\frac{SU(1,n)}{%
U(1)\otimes SU(n)}$ has $C_{ijk}=0$ globally ($n=n_{V}\in \mathbb{N}$). As
shown in App. I of \cite{bellucci1}, such a family has only two classes of
\textit{non-degenerate} solutions to the AEs: one $\frac{1}{2}$-BPS and one
non-BPS with $Z=0$.} $\frac{SU(1,n)}{U(1)\otimes SU(n)}$, \textit{all}
symmetric SK geometries are endowed with \textit{cubic} holomorphic
prepotentials. In rank-$3$ symmetric \textit{cubic} SK manifolds $\frac{G}{%
H=H_{0}\otimes U(1)}$ (which all are the vector supermultiplets' scalar
manifolds of $\mathcal{N}=2$, $d=4$ supergravities defined by \textit{Jordan
algebras} of degree $3$; see \textit{e.g.} \cite{bellucci1} and Refs.
therein), the solutions to AEs have been shown to exist in three distinct
classes, one $\frac{1}{2}$-BPS and the other two non-BPS, one of which
corresponds to vanishing central charge $Z=0$. It is here worth remarking
that the non-BPS $Z=0$ class of solutions to AEs has no analogue in $d=5$,
where a similar classification has been given \cite{FG2}.

Furthermore, the three classes of critical points of $V_{BH}$ in $\mathcal{N}%
=2$, $d=4$ symmetric cubic supergravities have been put in \textit{one-to-one%
} correspondence with the \textit{non-degenerate} charge orbits of the
actions of the $U$-duality groups $G$ on the corresponding BH charge
configuration spaces. In other words, the three species of solutions to AEs
in $\mathcal{N}=2$, $d=4$ symmetric cubic supergravities are supported by
configurations of the BH charges lying along the \textit{non-degenerate}
typologies of charge orbits of the $U$-duality group $G$ in the real
(electric-magnetic field strengths) representation space $R_{V}$,
determining its embedding in the symplectic group $Sp\left( 2n_{V}+2,\mathbb{%
R}\right) $. The results on charge orbits obtained in \cite{bellucci1} are
summarized\footnote{%
The charge orbits for the so-called $st^{2}$ and $stu$ models ($n=1$ and $%
n=2 $ elements of the cubic sequence $\frac{SU\left( 1,1\right) }{U\left(
1\right) }\otimes \frac{SO\left( 2,n\right) }{SO\left( 2\right) \otimes
SO\left( n\right) }$, respectively) are given in Appendix II of \cite
{bellucci1}, where also the charge orbits of the so-called $t^{3}$ model are
treated. It should be here pointed out that the $t^{3}$ model is an \textit{%
isolated case} in the classification of symmetric SK manifolds (see \textit{%
e.g.} \cite{CFG}), and it cannot be obtained as the $n=0$ element of the
cubic sequence $\frac{SU\left( 1,1\right) }{U\left( 1\right) }\otimes \frac{%
SO\left( 2,n\right) }{SO\left( 2\right) \otimes SO\left( n\right) }$, which
instead is the so-called $t^{2}$ model, given by the $n=1$ element of the
quadratic sequence $\frac{SU\left( 1,n\right) }{SU\left( n\right) \otimes
U\left( 1\right) }$, as well.} in Table 1.

\begin{table}[t]
\begin{center}
\begin{tabular}{|c||c|c|c|}
\hline
& $
\begin{array}{c}
\\
\frac{1}{2}\text{-BPS orbits } \\
~~\mathcal{O}_{\frac{1}{2}-BPS}=\frac{G}{H_{0}} \\
~
\end{array}
$ & $
\begin{array}{c}
\\
\text{non-BPS, }Z\neq 0\text{ orbits} \\
\mathcal{O}_{non-BPS,Z\neq 0}=\frac{G}{\widehat{H}}~ \\
~
\end{array}
$ & $
\begin{array}{c}
\\
\text{non-BPS, }Z=0\text{ orbits} \\
\mathcal{O}_{non-BPS,Z=0}=\frac{G}{\widetilde{H}}~ \\
~
\end{array}
$ \\ \hline\hline
$
\begin{array}{c}
\\
\mathit{Quadratic~Sequence} \\
(n=n_{V}\in \mathbb{N})
\end{array}
$ & $\frac{SU(1,n)}{SU(n)}~$ & $-$ & $\frac{SU(1,n)}{SU(1,n-1)}~$ \\ \hline
$
\begin{array}{c}
\\
\mathbb{R}\oplus \Gamma _{n} \\
(n=n_{V}-1\in \mathbb{N})
\end{array}
$ & $\frac{SU(1,1)\otimes SO(2,n)}{SO(2)\otimes SO(n)}~$ & $\frac{%
SU(1,1)\otimes SO(2,n)}{SO(1,1)\otimes SO(1,n-1)}~$ & $\frac{SU(1,1)\otimes
SO(2,n)}{SO(2)\otimes SO(2,n-2)}$ \\ \hline
$
\begin{array}{c}
\\
J_{3}^{\mathbb{O}} \\
~
\end{array}
$ & $\frac{E_{7(-25)}}{E_{6}}$ & $\frac{E_{7(-25)}}{E_{6(-26)}}$ & $\frac{%
E_{7(-25)}}{E_{6(-14)}}~$ \\ \hline
$
\begin{array}{c}
\\
J_{3}^{\mathbb{H}} \\
~
\end{array}
$ & $\frac{SO^{\ast }(12)}{SU(6)}~$ & $\frac{SO^{\ast }(12)}{SU^{\ast }(6)}~$
& $\frac{SO^{\ast }(12)}{SU(4,2)}~$ \\ \hline
$
\begin{array}{c}
\\
J_{3}^{\mathbb{C}} \\
~
\end{array}
$ & $\frac{SU(3,3)}{SU(3)\otimes SU(3)}$ & $\frac{SU(3,3)}{SL(3,\mathbb{C})}$
& $\frac{SU(3,3)}{SU(2,1)\otimes SU(1,2)}~$ \\ \hline
$
\begin{array}{c}
\\
J_{3}^{\mathbb{R}} \\
~
\end{array}
$ & $\frac{Sp(6,\mathbb{R})}{SU(3)}$ & $\frac{Sp(6,\mathbb{R})}{SL(3,\mathbb{%
R})}$ & $\frac{Sp(6,\mathbb{R})}{SU(2,1)}$ \\ \hline
\end{tabular}
\end{center}
\caption{\textbf{Non-degenerate charge orbits of the real, symplectic }$%
R_{V} $ \textbf{representation of the }$U$\textbf{-duality group }$G$
\textbf{supporting BH attractors with non-vanishing entropy in $\mathcal{N}%
=2 $, $d=4$ symmetric supergravities } \protect\cite{bellucci1}\textbf{\ }}
\end{table}

In all the $\mathcal{N}=2$, $d=4$ symmetric supergravities based on rank-$3$
SK cubic manifolds, the classical BH entropy is given by the \textit{%
Bekenstein-Hawking entropy-area formula} (\cite{BH1}; see also Eq. (\ref
{BHEA}))
\begin{equation}
S_{BH}=\frac{A_{H}}{4}=\pi \left. V_{BH}\right| _{\partial V_{BH}=0}=\pi
\sqrt{\left| \mathcal{I}_{4}\right| },
\end{equation}
where $\mathcal{I}_{4}$ is the (unique, \textit{quartic}\footnote{%
For the quadratic irreducible rank-$1$ infinite sequence $\frac{SU(1,n)}{%
U(1)\otimes SU(n)}$ the unique $G$-invariant is instead \textit{quadratic}
in the BH charges; it is positive for $\frac{1}{2}$-BPS orbits and negative
for the non-BPS ($Z=0$) ones (see App. I of \cite{bellucci1}).} in the BH
charges) \textit{moduli-independent} $G$-invariant built out of the
(considered non-degenerate charge orbit in the) representation $R_{V}$. $%
\frac{1}{2}$-BPS and non-BPS $Z=0$ classes have $\mathcal{I}_{4}>0$, while
the non-BPS $Z\neq 0$ class is characterized by $\mathcal{I}_{4}<0$.\bigskip

An interesting direction for further investigations concerns the study of
extremal BH attractors in more general, \textit{non-cubic} SK geometries. A
noteworthy example is given by the SK geometries of the scalar manifolds of
those $\mathcal{N}=2$, $d=4$ supergravities obtained as effective,
low-energy theories of $d=10$ Type IIB superstrings compactified on
Calabi-Yau threefolds ($CY_{3}$s), \textit{away from the limit of large
volume} of $CY_{3}$s.

Recently, \cite{BFMY} studied the extremal BH attractors in $n_{V}=1$\ SK
geometries obtained by compactifications (away from the limit of large
volume of the internal manifold) on a peculiar class of $CY_{3}$s, given by
the so-called (mirror) \textit{Fermat} $CY_{3}$s. Such threefolds are
classified by the \textit{Fermat parameter} $\frak{k}=5,6,8,10$, and they
were firstly found in \cite{Strom-Witten}. The fourth order linear
Picard-Fuchs (PF) ordinary differential Equations determining the
holomorphic fundamental period $4\times 1$\ vector for such a class of $1$%
-modulus $CY_{3}$s were found some time ago for $\frak{k}=5$\ in \cite
{CDLOGP1,CDLOGP2} (see also \cite{Cadavid-Ferrara,Ferrara-Lerche-1}), and
for $\frak{k}=6,8,10$\ in \cite{KT}.

More specifically, \cite{BFMY} dealt with\textbf{\ }the so-called \textit{%
Landau-Ginzburg} (LG) extremal BH attractors, \textit{i.e. }the solutions to
the AEs near the origin $z=0$\ (named \textit{LG point}) of the moduli space
$\mathcal{M}_{n_{V}=1}$ ($dim_{\mathbb{C}}\mathcal{M}_{n_{V}=1}=1$), and the
BH charge configurations supporting $z=0$ to be a critical point of $V_{BH}$
were explicitly determined, as well.

An intriguing development in such a framework would amount to extending to
the \textit{Fermat} $CY_{3}$-compactifications (away from the limit of large
volume of the threefold) the conjecture formulated in Sect. 5 of \cite
{K2-bis}. The conjecture was formulated in the framework of (the large
volume limit of $CY_{3}$-compactifications leading to) the remarkably
\textit{triality-symmetric} cubic $stu$ model \cite{BKRSW,Shmakova,K2-bis},
and it argues that the instability of the considered non-BPS ($Z\neq 0$)
critical points of $V_{BH}$ might be only \textit{apparent}, since such
attractors might correspond to \textit{multi-center stable attractor
solutions} (see also \textit{e.g.} \cite{Ferrara-Gimon,Gaiotto1,GLS1,Li} and
Refs. therein), whose stable nature should be \textit{``resolved''} only at
\textit{sufficiently small} distances. The extension of such a tempting
conjecture to non-BPS extremal BH LG attractors in \textit{Fermat} $CY_{3}$%
-compactifications would be interesting; in particular, the extension to the
non-BPS $Z=0$ case might lead to predict the existence (at least in the
considered peculiar $n_{V}=1$ framework) of \textit{non-BPS lines of
marginal stability} \cite{Denef1,Denef2} with $Z=0$.\smallskip

Moreover, it should be here recalled that the PF Eqs. of \textit{Fermat} $%
CY_{3}$s (\cite{CDLOGP1}-\nocite{CDLOGP2, Cadavid-Ferrara}\cite{KT}, see
also \cite{BFMY}) exhibit other two species of \textit{regular singular}
points, namely the $k$-th roots of unity ($z^{k}=1$, the so-called \textit{%
conifold points}) and the \textit{point at infinity} $z\longrightarrow
\infty $ in the moduli space, corresponding to the so-called \textit{large
complex structure modulus limit}. Thus, it would be interesting to solve the
AEs in proximity of such regular singular points, \textit{i.e.} it would be
worth investigating \textit{extremal BH conifold attractors} and \textit{%
extremal BH large complex structure attractors} in the moduli space of
1-modulus (\textit{Fermat}) $CY_{3}$s. Such an investigation would be of
interest, also in view of recent studies of extremal BH attractors in
peculiar examples of $n_{V}=2$-moduli $CY_{3}$-compactifications \cite
{Misra1}.\bigskip

Let us now consider the crucial issue of stability more in detail.

In $\mathcal{N}=2$ homogeneous (not necessarily symmetric) and $\mathcal{N}%
>2 $-extended (all symmetric), $d=4$ supergravities the Hessian matrix of $%
V_{BH}$ at its critical points is in general \textit{semi-positive definite}%
, eventually with some vanishing eigenvalues (\textit{massless Hessian modes}%
), which actually are \textit{flat} directions of $V_{BH}$ itself \cite
{Ferrara-Marrani-1,ferrara4}. Thus, it can be stated that for all
supergravities based on homogeneous scalar manifolds the critical points of $%
V_{BH}$ which are \textit{non-degenerate} (\textit{i.e.} for which it holds $%
V_{BH}\neq 0$) all are \textit{stable}, up to some eventual \textit{flat}
directions.

As pointed out above, the Attractor Equations of $\mathcal{N}=2$, $d=4$
supergravity with $n_{V}$ Abelian vector multiplets may have \textit{flat}
directions in the non-BPS cases \cite{Ferrara-Marrani-1,ferrara4}, but
\textit{not} in the $\frac{1}{2}$-BPS one \cite{FGK}. Indeed, in the $\frac{1%
}{2}$-BPS case (satisfying $Z\neq 0$, $D_{i}Z=0$ $\forall i=1,...,n_{V}$;
recall Eq. (\ref{BPS-conds})) the covariant $2n_{V}\times 2n_{V}$ Hessian
matrix of $V_{BH}$ reads (\cite{FGK}; recall Eqs. (\ref{SUSY-crit}))
\begin{equation}
\left( D_{\widehat{i}}D_{\widehat{j}}V_{BH}\right) _{\mathcal{N}=2,\frac{1}{2%
}-BPS}=\frac{1}{2}\left| Z\right| _{\frac{1}{2}-BPS}\left(
\begin{array}{ccc}
0 &  & g_{i\overline{j}} \\
&  &  \\
g_{j\overline{i}} &  & 0
\end{array}
\right) _{\frac{1}{2}-BPS},  \label{b-sunday}
\end{equation}
where hatted indices can be either holomorphic or anti-holomorphic; thus, as
far as the metric $g_{i\overline{j}}$ of the scalar manifold is strictly
positive definite, Eq. (\ref{b-sunday}) yields that no \textit{massless} $%
\frac{1}{2}$-BPS \textit{Hessian modes} arise out.
\begin{table}[t]
\begin{center}
\begin{tabular}{|c||c|c|c|}
\hline
& $
\begin{array}{c}
\\
\frac{\widehat{H}}{\widehat{h}} \\
~
\end{array}
$ & $
\begin{array}{c}
\\
\text{r} \\
~
\end{array}
$ & $
\begin{array}{c}
\\
\text{\textit{dim}}_{\mathbb{R}} \\
~
\end{array}
$ \\ \hline\hline
$
\begin{array}{c}
\\
\mathbb{R}\oplus \Gamma _{n} \\
(n=n_{V}-1\in \mathbb{N})
\end{array}
$ & $SO(1,1)\otimes \frac{SO(1,n-1)}{SO(n-1)}~$ & $
\begin{array}{c}
\\
1(n=1) \\
2(n\geqslant 2) \\
~
\end{array}
$ & $n~$ \\ \hline
$
\begin{array}{c}
\\
J_{3}^{\mathbb{O}} \\
~
\end{array}
$ & $\frac{E_{6(-26)}}{F_{4(-52)}}~$ & $2~$ & $6~$ \\ \hline
$
\begin{array}{c}
\\
J_{3}^{\mathbb{H}} \\
~
\end{array}
$ & $\frac{SU^{\ast }(6)}{USp(6)}~$ & $2~$ & $14~$ \\ \hline
$
\begin{array}{c}
\\
J_{3}^{\mathbb{C}} \\
~
\end{array}
$ & $\frac{SL(3,C)}{SU(3)}~$ & $2~$ & $8~$ \\ \hline
$
\begin{array}{c}
\\
J_{3}^{\mathbb{R}} \\
~
\end{array}
$ & $\frac{SL(3,\mathbb{R})}{SO(3)}~$ & $2~$ & $5$ \\ \hline
\end{tabular}
\end{center}
\caption{\textbf{Moduli spaces of non-degenerate non-BPS }$Z\neq 0$ \textbf{%
\ critical points of } $V_{BH,\mathcal{N}=2}$ \textbf{in }$\mathcal{N}=2,d=4$
\textbf{symmetric supergravities (}$\widehat{h}$\textbf{\ is the maximal
compact subgroup of }$\widehat{H}$).\textbf{\ They are the} $\mathcal{N}%
=2,d=5$ \textbf{symmetric real special manifolds} \protect\cite{ferrara4} }
\end{table}

\begin{table}[h]
\begin{center}
\begin{tabular}{|c||c|c|c|}
\hline
& $
\begin{array}{c}
\\
\frac{\widetilde{H}}{\widetilde{h}}=\frac{\widetilde{H}}{\widetilde{h}%
^{\prime }\otimes U(1)} \\
~
\end{array}
$ & $
\begin{array}{c}
\\
\text{r} \\
~
\end{array}
$ & $
\begin{array}{c}
\\
\text{\textit{dim}}_{\mathbb{C}} \\
~
\end{array}
$ \\ \hline\hline
$
\begin{array}{c}
\\
Quadratic~~Sequence \\
(n=n_{V}\in \mathbb{N}) \\
~
\end{array}
$ & $\frac{SU(1,n-1)}{U(1)\otimes SU(n-1)}~$ & $1~$ & $n-1~$ \\ \hline
$
\begin{array}{c}
\\
\mathbb{R}\oplus \Gamma _{n} \\
(n=n_{V}-1\in \mathbb{N})
\end{array}
$ & $\frac{SO(2,n-2)}{SO(2)\otimes SO(n-2)},n\geqslant 3~$ & $
\begin{array}{c}
\\
1(n=3) \\
2(n\geqslant 4) \\
~
\end{array}
$ & $n-2~$ \\ \hline
$
\begin{array}{c}
\\
J_{3}^{\mathbb{O}} \\
~
\end{array}
$ & $\frac{E_{6(-14)}}{SO(10)\otimes U(1)}~$ & $2~$ & $16~$ \\ \hline
$
\begin{array}{c}
\\
J_{3}^{\mathbb{H}} \\
~
\end{array}
$ & $\frac{SU(4,2)}{SU(4)\otimes SU(2)\otimes U(1)}~$ & $2~$ & $8~$ \\ \hline
$
\begin{array}{c}
\\
J_{3}^{\mathbb{C}} \\
~
\end{array}
$ & $\frac{SU(2,1)}{SU(2)\otimes U(1)}\otimes \frac{SU(1,2)}{SU(2)\otimes
U(1)}~$ & $2~$ & $4~$ \\ \hline
$
\begin{array}{c}
\\
J_{3}^{\mathbb{R}} \\
~
\end{array}
$ & $\frac{SU(2,1)}{SU(2)\otimes U(1)}~$ & $1~$ & $2~$ \\ \hline
\end{tabular}
\end{center}
\caption{\textbf{Moduli spaces of non-degenerate non-BPS }$Z=0$ \textbf{\
critical points of } $V_{BH,\mathcal{N}=2}$ \textbf{in }$\mathcal{N}=2,d=4$
\textbf{symmetric supergravities (}$\widetilde{h}$\textbf{\ is the maximal
compact subgroup of }$\widetilde{H}$)\textbf{. They are (non-special)
symmetric K\"{a}hler manifolds} \protect\cite{ferrara4}}
\end{table}

Tables 2 and 3 respectively list the moduli spaces of non-BPS $Z\neq 0$ and
non-BPS $Z=0$ attractors for symmetric $\mathcal{N}=2$, $d=4$ SK geometries,
for which a complete classification is available \cite{ferrara4} (the
attractor moduli spaces should exist also in homogeneous non-symmetric $%
\mathcal{N}=2$, $d=4$\ SK geometries, but their classification is currently
unknown). The general \textit{thumb rule} to construct the moduli space of a
given attractor solution in the considered symmetric framework is to coset
the \textit{stabilizer} of the corresponding charge orbit by its \textit{%
maximal compact subgroup}. By such a rule, the $\frac{1}{2}$-BPS attractors
do \textit{not} have an associated moduli space simply because the
stabilizer of their supporting BH charge orbit is \textit{compact}. On the
other hand, \textit{all} attractors supported by BH charge orbits whose
stabilizer is \textit{non-compact} exhibit a non-vanishing moduli space.
furthermore, it should be noticed that the non-BPS $Z\neq 0$ moduli spaces
are nothing but the symmetric real special scalar manifolds of the
corresponding $\mathcal{N}=2$, $d=5$ supergravity.\smallskip

Nevertheless, it is worth remarking that some symmetric $\mathcal{N}=2$, $%
d=4 $ supergravities have no non-BPS \textit{flat }directions at all.

The unique $n_{V}=1$ symmetric models are the so-called $t^{2}$ and $t^{3}$
models; they are based on the rank-$1$ scalar manifold $\frac{SU\left(
1,1\right) }{U\left( 1\right) }$, but with different holomorphic
prepotential functions. The $t^{2}$ model is the first element ($n=1$) of
the sequence of irreducible symmetric special K\"{a}hler manifolds $\frac{%
SU\left( 1,n\right) }{U\left( 1\right) \times SU\left( n\right) }$ ($n_{V}=n$%
, $n\in \mathbb{N}$) (see \textit{e.g.} \cite{bellucci1} and Refs. therein),
endowed with \textit{quadratic} prepotential. Its bosonic sector is given by
the $\left( U\left( 1\right) \right) ^{6}\rightarrow \left( U\left( 1\right)
\right) ^{2}$ truncation of Maxwell-Einstein-axion-dilaton (super)gravity,
\textit{i.e.} of \textit{pure} $\mathcal{N}=4$, $d=4$ supergravity. On the
other hand, the $t^{3}$ model has \textit{cubic} prepotential; as pointed
out above, it is an \textit{isolated case} in the classification of
symmetric SK manifolds (see \textit{e.g.} \cite{CFG}), but it can be thought
also as the $\mathit{s=t=u}$\textit{\ degeneration }of the $stu$ model. It
is worth pointing out that the $t^{2}$ and $t^{3}$ models are based on the
same rank-$1$ SK manifold, with different constant \textit{scalar curvature}%
, which respectively can be computed to be (see \textit{e.g.} \cite
{BFM-SIGRAV06} and Refs. therein)
\begin{equation}
\begin{array}{l}
\frac{SU(1,1)}{U(1)},~t^{2}~\text{\textit{model}}:R=-2; \\
\\
\frac{SU(1,1)}{U(1)},\text{~}t^{3}~\text{\textit{model}}:R=-\frac{2}{3}.
\end{array}
\end{equation}
Beside the $\frac{1}{2}$-BPS attractors, the $t^{2}$ model admits only
non-BPS $Z=0$ critical points of $V_{BH}$ with no \textit{flat} directions.
Analogously, the $t^{3}$ model admits only non-BPS $Z\neq 0$ critical points
of $V_{BH}$ with no \textit{flat} directions.

For $n_{V}>1$, the non-BPS $Z\neq 0$ critical points of $V_{BH}$, if any,
all have \textit{flat} directions, and thus a related moduli space (see
Table 1). However, models with no non-BPS $Z=0$ \textit{flat} directions at
all and $n_{V}>1$ exist, namely they are the first and second element ($n=1$%
, $2$) of the sequence of reducible symmetric special K\"{a}hler manifolds $%
\frac{SU\left( 1,1\right) }{U\left( 1\right) }\times \frac{SO\left(
2,n\right) }{SO\left( 2\right) \times SO\left( n\right) }$ ($n_{V}=n+1$, $%
n\in \mathbb{N}$) (see \textit{e.g.} \cite{bellucci1} and Refs. therein),
\textit{i.e.} the so-called $st^{2}$ and $stu$ models, respectively. The $%
stu $ model (relevant also for the recently established connection between
extremal BHs and \textit{Quantum Information Theory} \cite{Duff}--\nocite
{KL,Levay,Ferrara-Duff1,Levay2}\cite{Ferrara-Duff2}) has two non-BPS $Z\neq
0 $ \textit{flat} directions, spanning the moduli space $SO\left( 1,1\right)
\times SO\left( 1,1\right) $ (\textit{i.e.} the scalar manifold of the $stu$
model in $d=5$), but \textit{no} non-BPS $Z=0$ \textit{massless Hessian modes%
} at all. On the other hand, the $st^{2}$ model (which can be thought as the
$\mathit{t=u}$\textit{\ degeneration} of the $stu$ model) has one non-BPS $%
Z\neq 0$ \textit{flat} direction, spanning the moduli space $SO\left(
1,1\right) $ (\textit{i.e.} the scalar manifold of the $st^{2}$ model in $%
d=5 $), but \textit{no} non-BPS $Z=0$ \textit{flat} direction at all. The $%
st^{2} $ is the \textit{``smallest''} symmetric model exhibiting a non-BPS $%
Z\neq 0$ \textit{flat} direction.

Concerning the \textit{``smallest''} symmetric models exhibiting a non-BPS $%
Z=0$ \textit{flat} direction they are the second ($n=2$) element of the
sequence $\frac{SU\left( 1,n\right) }{U\left( 1\right) \times SU\left(
n\right) }$ and the third ($n=3$) element of the sequence $\frac{SU\left(
1,1\right) }{U\left( 1\right) }\times \frac{SO\left( 2,n\right) }{SO\left(
2\right) \times SO\left( n\right) }$. In both cases, the unique non-BPS $Z=0$
\textit{flat} direction spans the non-BPS $Z=0$ moduli space $\frac{SU\left(
1,1\right) }{U\left( 1\right) }\sim \frac{SO\left( 2,1\right) }{SO\left(
2\right) }$ (see Table 2), whose local geometrical properties however differ
in the two cases (for the same reasons holding for the $t^{2}$ and $t^{3}$
models treated above).

\begin{table}[t]
\begin{center}
\begin{tabular}{|c||c|c|c|}
\hline
& $
\begin{array}{c}
\\
\frac{1}{\mathcal{N}}\text{-BPS orbits } \frac{G}{\mathcal{H}} \\
~
\end{array}
$ & $
\begin{array}{c}
\\
\text{non-BPS, }Z_{AB}\neq 0\text{ orbits}~\frac{G}{\widehat{\mathcal{H}}}
\\
~
\end{array}
$ & $
\begin{array}{c}
\\
\text{non-BPS, }Z_{AB}=0\text{ orbits }\frac{G}{\widetilde{\mathcal{H}}} \\
\\
~
\end{array}
$ \\ \hline\hline
$
\begin{array}{c}
\\
\mathcal{N}=3 \\
~
\end{array}
$ & $\frac{SU(3,n)}{SU(2,n)}~$ & $-$ & $\frac{SU(3,n)}{SU(3,n-1)}~$ \\ \hline
$
\begin{array}{c}
\\
\mathcal{N}=4 \\
~
\end{array}
$ & $\frac{SU(1,1)}{U(1)}\otimes \frac{SO(6,n)}{SO(4,n)}~$ & $\frac{SU(1,1)}{%
SO(1,1)}\otimes \frac{SO(6,n)}{SO(5,n-1)}~$ & $\frac{SU(1,1)}{U(1)}\otimes
\frac{SO(6,n)}{SO(6,n-2)}$ \\ \hline
$
\begin{array}{c}
\\
\mathcal{N}=5 \\
~
\end{array}
$ & $\frac{SU(1,5)}{SU(3)\otimes SU\left( 2,1\right) }$ & $-$ & $-$ \\ \hline
$
\begin{array}{c}
\\
\mathcal{N}=6 \\
~
\end{array}
$ & $\frac{SO^{\ast }(12)}{SU(4,2)}~$ & $\frac{SO^{\ast }(12)}{SU^{\ast }(6)}%
~$ & $\frac{SO^{\ast }(12)}{SU(6)}~$ \\ \hline
$
\begin{array}{c}
\\
\mathcal{N}=8 \\
~
\end{array}
$ & $\frac{E_{7\left( 7\right) }}{E_{6\left( 2\right) }}$ & $\frac{%
E_{7\left( 7\right) }}{E_{6\left( 6\right) }}$ & $-~$ \\ \hline
\end{tabular}
\end{center}
\caption{\textbf{Non-degenerate charge orbits of the real, symplectic }$%
R_{V} $ \textbf{representation of the }$U$\textbf{-duality group }$G$
\textbf{supporting BH attractors with non-vanishing entropy in $\mathcal{N}%
>2 $-extended, $d=4$ supergravities} \textbf{(}$n$\textbf{\ is the number of
matter multiplets) }\protect\cite{review-Kallosh}}
\end{table}

\begin{table}[t]
\begin{center}
\begin{tabular}{|c||c|c|c|}
\hline
& $
\begin{array}{c}
\\
\frac{1}{\mathcal{N}}\text{-BPS} \\
\text{moduli space }\frac{\mathcal{H}}{\frak{h}}\text{ } \\
~
\end{array}
$ & $
\begin{array}{c}
\\
\text{non-BPS, }Z_{AB}\neq 0 \\
\text{moduli space }\frac{\widehat{\mathcal{H}}}{\widehat{\frak{h}}} \\
~
\end{array}
$ & $
\begin{array}{c}
\\
\text{non-BPS, }Z_{AB}=0 \\
\text{moduli space }\frac{\widetilde{\mathcal{H}}}{\widetilde{\frak{h}}} \\
~
\end{array}
$ \\ \hline\hline
$
\begin{array}{c}
\\
\mathcal{N}=3 \\
~
\end{array}
$ & $\frac{SU(2,n)}{SU(2)\otimes SU\left( n\right) \otimes U\left( 1\right) }%
~$ & $-$ & $\frac{SU(3,n-1)}{SU(3)\otimes SU\left( n-1\right) \otimes
U\left( 1\right) }~$ \\ \hline
$
\begin{array}{c}
\\
\mathcal{N}=4 \\
~
\end{array}
$ & $\frac{SO(4,n)}{SO(4)\otimes SO\left( n\right) }~$ & $SO(1,1)\otimes
\frac{SO(5,n-1)}{SO(5)\otimes SO\left( n-1\right) }~$ & $\frac{SO(6,n-2)}{%
SO(6)\otimes SO\left( n-2\right) }$ \\ \hline
$
\begin{array}{c}
\\
\mathcal{N}=5 \\
~
\end{array}
$ & $\frac{SU\left( 2,1\right) }{SU\left( 2\right) \otimes U\left( 1\right) }
$ & $-$ & $-$ \\ \hline
$
\begin{array}{c}
\\
\mathcal{N}=6 \\
~
\end{array}
$ & $\frac{SU(4,2)}{SU(4)\otimes SU\left( 2\right) \otimes U\left( 1\right) }%
~$ & $\frac{SU^{\ast }(6)}{USp\left( 6\right) }~$ & $-$ \\ \hline
$
\begin{array}{c}
\\
\mathcal{N}=8 \\
~
\end{array}
$ & $\frac{E_{6\left( 2\right) }}{SU\left( 6\right) \otimes SU\left(
2\right) }$ & $\frac{E_{6\left( 6\right) }}{USp\left( 8\right) }$ & $-~$ \\
\hline
\end{tabular}
\end{center}
\caption{\textbf{Moduli spaces of BH attractors with non-vanishing entropy
in $\mathcal{N}>2$-extended, $d=4$ supergravities (}$\frak{h}$\textbf{, }$%
\widehat{\frak{h}}$\textbf{\ and }$\widetilde{\frak{h}}$\textbf{\ are
maximal compact subgroups of }$\mathcal{H}$\textbf{, }$\widehat{\mathcal{H}}$%
\textbf{\ and }$\widetilde{\mathcal{H}}$\textbf{, respectively, and }$n$
\textbf{is the number of matter multiplets)} \protect\cite{review-Kallosh}}
\end{table}

\subsection{\label{N>2,d=4}$\mathcal{N}>2$-Extended, $d=4$ Supergravity}

In $\mathcal{N}>2$-extended, $d=4$ supergravities, whose scalar manifold is
always symmetric, there are \textit{flat} directions of $V_{BH}$ at both its
\textit{non-degenerate} BPS and non-BPS critical points. As mentioned above,
from a group-theoretical point of view this is due to the fact that the
corresponding supporting BH charge orbits always have a \textit{non-compact}
stabilizer \cite{ferrara4,review-Kallosh}. The BPS \textit{flat} directions
can be interpreted in terms of left-over hypermultiplets' scalar degrees of
freedom in the truncation down to the $\mathcal{N}=2$, $d=4$ theories \cite
{ADF-U-duality-d=4,Ferrara-Marrani-1}. In Tables 4 and 5 all charge orbits
and the corresponding moduli spaces of attractor solution in $\mathcal{N}>2$%
-extended, $d=4$ supergravities are reported \cite{review-Kallosh}.

We conclude by pointing out that in the present report we dealt with results
holding at the classical, Einstein supergravity level. It is conceivable
that the \textit{flat} directions of classical \textit{non-degenerate}
extremal BH attractors will be removed (\textit{i.e.} lifted) by \textit{%
quantum} (\textit{perturbative} and \textit{non-perturbative}) corrections
(such as the ones coming from higher-order derivative contributions to the
gravity and/or gauge sector) to the \textit{classical} effective BH
potential $V_{BH}$. Consequently, \textit{at the quantum} (\textit{%
perturbative} and \textit{non-perturbative}) \textit{level, no moduli spaces
for attractor solutions might exist at all} (and therefore also \textit{the
actual attractive nature of the critical points of }$V_{BH}$\textit{\ might
be destroyed}). \textit{However, this might not be the case for }$\mathcal{N}%
=8$.

In presence of \textit{quantum} lifts of \textit{classically flat}
directions of the Hessian matrix of $V_{BH}$ at its critical points, in
order to answer to the key question: \textit{``Do extremal BH attractors (in
a strict sense) survive the quantum level?''}, it is thus crucial to
determine whether such lifts originate Hessian modes with \textit{positive}
squared mass (corresponding to \textit{attractive} directions) or with
\textit{negative} squared mass (\textit{i.e.} \textit{tachyonic}, \textit{%
repeller} directions).

The fate of the unique non-BPS $Z\neq 0$ flat direction of the $st^{2}$
model in presence of the most general class of quantum perturbative
corrections consistent with the axionic-shift symmetry has been studied in
\cite{BFMS2}, showing that, as intuitively expected, the \textit{classical
solutions get lifted at the quantum level}. Interestingly, in \cite{BFMS2}%
\textbf{\ }it is found the \textit{quantum} lift occurs more often towards
\textit{repeller} directions (thus destabilizing the whole critical
solution, and \textit{destroying the attractor in strict sense}), rather
than towards \textit{attractive} directions.\textbf{\ }The same behavior may
be expected for the unique non-BPS $Z=0$\ flat direction of the $n=2$\
element of the quadratic irreducible sequence and the $n=3$\ element of the
cubic reducible sequence (see above).

Generalizing to the presence of more than one \textit{flat} direction, this
would mean that \textit{only a (very) few classical attractors do remain
attractors in strict sense at the quantum level}; consequently, \textit{at
the quantum} (\textit{perturbative} and \textit{non-perturbative}) \textit{%
level the ``landscape'' of extremal BH attractors should be strongly
constrained and reduced}.\medskip

Despite the considerable number of papers written on the \textit{Attractor
Mechanism} in the extremal BHs of the supersymmetric theories of gravitation
along the last years, still much remains to be discovered along the way
leading to a deep understanding of the inner dynamics of (eventually
extended) space-time singularities in supergravities, and hopefully in their
fundamental high-energy counterparts, such as $d=10$ superstrings and $d=11$
$M$-theory.

\section*{\textbf{Acknowledgments}}

The original parts of the contents of this report result from collaborations
with L. Andrianopoli, A. Ceresole, R. D'Auria, E. Gimon, M. Trigiante, and
especially G. Gibbons, M. G\"{u}naydin, R. Kallosh and A. Strominger, which
are gratefully acknowledged.

A. M. would like to thank the Department of Physics, Theory Unit Group at
CERN, where part of this work was done, for kind hospitality and stimulating
environment.

The work of S.B. has been supported in part by the European Community Human
Potential Program under contract MRTN-CT-2004-005104 \textit{``Constituents,
Fundamental Forces and Symmetries of the Universe''}.

The work of S.F. has been supported in part by European Community Human
Potential Program under contract MRTN-CT-2004-005104 \textit{``Constituents,
Fundamental Forces and Symmetries of the Universe''}, in association with
INFN Frascati National Laboratories and by D.O.E. grant DE-FG03-91ER40662,
Task C.

The work of A.M. has been supported by a Junior Grant of the \textit{%
``Enrico Fermi''} Center, Rome, in association with INFN Frascati National
Laboratories.

\end{document}